\documentclass[a4paper,11pt]{article}
\pdfoutput=1 

\usepackage{jcappub} 
\usepackage{xcolor}

\usepackage[T1]{fontenc} 

\title{\boldmath Game Theory in Cosmology}

\author[1]{Oem Trivedi,}
\author[2]{Venkat Venkatasubramanian}

\affiliation[1]{Department of Physics and Astronomy, Vanderbilt University, Nashville, TN 37235, USA}
\affiliation[2]{Complex Resilient Intelligent Systems Laboratory, Department of Chemical Engineering, Columbia University, New York, NY 10027, U.S.A.}
\emailAdd{oem.trivedi@vanderbilt.edu}
\emailAdd{venkat@columbia.edu}

\abstract{We present a game-theoretic statistical framework for cosmology, which we term \textit{Cosmic Teleodynamics}. We recast the dark sector, cosmic acceleration, large-scale structure, and cosmic tensions as emergent consequences of nonlocal memory and intrinsically persistent organization in a self-gravitating universe. Our main insight is that, just as the Big Bang and Inflation left their signature on the Cosmic Microwave Background (CMB), they also left a similar lasting imprint on the spacetime itself, influencing future dynamics and evolution. By introducing a maximum-caliber weight on cosmic histories and a bias functional encoding structural memory, we derive modified Friedmann, Boltzmann, and Poisson equations that naturally generate dark energy-like acceleration, dark matter-like clustering, and scale-dependent growth suppression. We also show how this approach can naturally help alleviate the $H_0$ and $S_8$ tensions, can produce anisotropic velocity fields, and predict environment-dependent halo signatures that cannot arise from particle dark matter or scalar-field dark energy. We also derive a generalized horizon entropy and temperature, revealing a nonequilibrium statistical origin for cosmic acceleration. We formulate a Law of Universal Arbitrage Equilibrium that governs the evolution of the Universe, showing that it is expanding towards a continuous form of Nash equilibrium. We also show that our approach naturally explains the Coincidence problem as well. Cosmic Teleodynamics therefore offers a unified, emergent, and testable alternative to the conventional dark sector, rooted not in new particles but in the intrinsic statistical and systemic structure of cosmic memory. The crux of this framework is, by just considering a more complete form of statistical mechanics, one can address multiple major problems in modern cosmology.}

\begin{document}
\maketitle
\flushbottom

\section{Introduction}
Modern cosmology has achieved remarkable success in describing the evolution of the universe from the Big Bang to the present day, yet it faces profound questions that challenge our understanding of fundamental physics. The simplest and most successful model of cosmology today, the $\Lambda$CDM model incorporating cold dark matter (CDM) and a cosmological constant ($\Lambda$), provides an elegant framework that matches numerous observations, from the cosmic microwave background (CMB) to the large-scale distribution of galaxies. However, this success comes at a fundamental cost, as approximately 95\% of the universe's energy budget consists of dark components, namely, dark energy \cite{de1SupernovaSearchTeam:1998fmf} and dark matter \cite{dm11rubin1970rotation}, whose fundamental nature remains unknown. The explanations offered for dark matter range from primordial black holes to axions \cite{dm1Cirelli:2024ssz,dm2Arbey:2021gdg,dm3Balazs:2024uyj,dm4Eberhardt:2025caq,dm5Bozorgnia:2024pwk,dm6Misiaszek:2023sxe,dm7OHare:2024nmr,dm8Adhikari:2022sbh,dm9Miller:2025yyx,dm10Trivedi:2025vry,dm12Trivedi:2025sbe}, while the ones for dark energy range from the simple cosmological constant to a wide array of exotic scalar field models and modified gravity \cite{de2Li:2012dt,de3Li:2011sd,de4Mortonson:2013zfa,de5Frusciante:2019xia,de6Huterer:2017buf,de7Vagnozzi:2021quy,de8Adil:2023ara,de9Feleppa:2025clx,de10DiValentino:2020evt,de11Nojiri:2010wj,de12Nojiri:2006ri,de13Trivedi:2023zlf,de14Trivedi:2022svr,de15Trivedi:2024inb}. Furthermore, in later-stage cosmology, one confronts observational discrepancies over the exact values of the universe's most fundamental parameters, such as the Hubble Constant ($H_0$) and the Sigma 8 ($\sigma_8$) tensions \cite{ht1DiValentino:2021izs,ht2Clifton:2024mdy,s81kazantzidis2018evolution,s82amon2022non,s83poulin2023sigma,s84Ferreira:2025lrd}. \\

Beyond these component mysteries lies a subtler class of phenomena involving gravitational memory and dynamically induced effects. \textit{Persistence} effects in large-scale structure refer to how cosmic evolution retains information about initial conditions longer than ideal models assume. The web-like structure of the universe — with its filaments, walls, and voids — encodes a remarkably detailed fossil record of primordial density fluctuations and subsequent gravitational dynamics. Gravitational memory effects and dynamically induced preferences emerge when gravitational interactions select particular configurations from initially Boltzmann-weighted distributions. Such alignment effects shape cosmic structure and may constrain modified gravity theories or dark matter properties beyond those accessible through traditional clustering statistics. These subtle phenomena of persistence exemplify how gravity's cumulative effects give rise to emergent structures that encode rich information about fundamental physics. \\

To address these concerns, we present a perspective that accounts for persistence effects in a novel manner by using game theory.  One might wonder how game theory applies to cosmology. It may seem rather far-fetched. However, as we demonstrate, there is a surprisingly close connection via the concept of persistence. Game theory is concerned with predicting the behavior of \textit{goal-driven} agents that, by their very nature of incessantly pursuing goals, exhibit persistent behavior. This persistence results in constrained behavior, where persistent agents do not explore the entire phase space but only a subset of it. Thus, persistence implies \textit{latent constraints} on the dynamics and evolution of agents. Game theory provides a new approach to modeling persistence mathematically as constraints on possible collective behaviors. It is in this sense that we use the word `game' in our theory of cosmology. \\

In particular, we use a novel game-theoretic framework, \textit{statistical teleodynamics}, originally developed to model the collective behavior of millions of goal-driven agents in biology, ecology, economics, and sociology~\cite{Venkat2025activematter,venkat2017book,venkat2025social}. Statistical teleodynamics is a synthesis of the fundamental principles of the theory of potential games~\cite{monderer1996potential, sandholm2010population} and those of statistical mechanics. The name originates from the Greek word \textit{telos}, meaning `goal'. Just as the dynamical behavior of gas molecules in a container is driven by thermal agitation (hence \textit{thermo}dynamics), the dynamics of active agents is driven by their persistent pursuit of goals, and hence \textit{teleo}dynamics. This theory applies to three kinds of goal-driven agents: (i) \textit{intentionally} persistent (e.g., humans), (ii) \textit{instinctively} persistent (e.g., bacteria, ants, birds), and (iii) \textit{intrinsically} persistent (e.g., active colloids, galaxies). In the limiting case of goal-free entities, such as gas molecules in a container, this theory reduces to the familiar results of statistical mechanics. \\

Thus, we present a teleodynamic model of the universe, which we call \textit{Cosmic Teleodynamics}, whose evolution is not purely random, but is influenced by persistent structures, memory effects, and dynamically induced preferences. In conventional statistical mechanics, one assumes that the probability assigned to a microscopic configuration is determined solely by its energy via the Boltzmann weight. One further assumes ergodicity, loss of memory, short-range interactions, etc. These assumptions are questionable for gravitational and cosmological systems where correlations are long-ranged, ergodicity is broken, and macroscopic evolution depends sensitively on prior structure formation. In this sense, teleodynamics expands the statistical toolkit by adding a second structural ingredient, in addition to energy, namely a functional encoding of the system's persistent organizational bias.

\section{Potential Games and Arbitrage Equilibrium}

Before we demonstrate the use of game-theoretic concepts in cosmology, we need to take a brief detour to introduce potential games~\cite{sandholm2010population}. The theory of potential games focuses on predicting the eventual outcome(s) of a dynamical system consisting of many competing, goal-driven agents, as observed in biology, economics, and sociology. For these games, one can identify a single scalar-valued global function, called a {\em potential} ($\phi(\boldsymbol{x})$), which captures the necessary information about the utilities {(where $\boldsymbol{x}$ is the state vector of the system)}. Here, agents continuously pursue arbitrage opportunities to increase their benefits and minimize costs in a competitive environment, thereby optimizing their benefit-cost trade-offs, which we define as \textit{effective utility} in our theory. The effective utility is essentially a proxy for persistence. Agents persistently seek arbitrage opportunities to increase their effective utility. This collective persistence dynamics results in an equilibrium, known as the Nash equilibrium, when the teleodynamic potential $\phi$ is maximized~\cite{sandholm2010population}. \\

The effective utility of an agent, $h_j$, in state $j$ is the gradient of potential $\phi(\boldsymbol{x})$, i.e.,
\begin{equation}
{h}_j(\boldsymbol{x})\equiv {\partial \phi(\boldsymbol{x})}/{\partial x_j}
\label{eq:utility-grad-phi}
\end{equation}
where $x_j=N_j/N$ and $\boldsymbol{x}$ is the population vector. {$N_j$ is the number of agents in the state $j$, and $N$ is the total number of agents}. Therefore, we have the following: 
\begin{eqnarray}
\phi(\boldsymbol{x})&=&\sum_{j=1}^n\int {h}_j(\boldsymbol{x}){d}x_j 
\label{eq:potential-defn}
\end{eqnarray}
where $n$ is the total number of states. To determine the maximum teleodynamic potential, we use the method of Lagrange multipliers with $\mathcal{L}$ as the Lagrangian and $\lambda$ as the Lagrange multiplier for the constraint $\sum_{j=1}^nx_j=1$:
\begin{equation}
\mathcal{L}=\phi+\lambda(1-\sum_{j=1}^nx_j)
\label{eq:lagrangian}
\end{equation}
In equilibrium, all agents enjoy the same utility. That is,
\begin{equation}
h_j = h^*    
\label{eq:hj=h*}
\end{equation}
It is an \textit{arbitrage equilibrium} \cite{kanbur2020occupational} in which agents no longer have any incentive to switch states as all states provide the same utility $h^*$. Thus, the maximization of $\phi$ and $h_j = h^*$ is equivalent when the equilibrium is unique (i.e., $\phi(\boldsymbol{x})$ is strictly concave \cite{sandholm2010population}). \\

One can view the thermodynamics of gas molecules in a container via the lens of potential games - the Thermodynamic Game~\cite{venkat2015howmuch}- and define the effective ``utility" $h_j$ of a molecule in state $j$ as 
\begin{equation}
h_{j}= -\beta E_j - \ln N_j
\label{eq:utility-thermo}
\end{equation}
where $E_j$ is its energy, $N_j$ is the number of molecules in state $j$, and $\beta = 1/k_BT$ is the Boltzmann factor. Using Eq.~\ref{eq:utility-thermo} in Eq.~\ref{eq:potential-defn}, we have
\begin{equation}
\phi(\textbf{x}) = -\beta \sum_{j=1}^{n} x_j E_j + \frac{1}{N} \ln \frac{N!}{\prod_{j=1}^{n} (N_j)!}
\label{eq:potential-thermo}
\end{equation}
Using Eq.~\ref{eq:potential-thermo} in Eq.~\ref{eq:lagrangian}, we obtain the well-known Gibbs-Boltzmann exponential probability distribution at equilibrium as
\begin{equation}
x_j^* = \frac{e^{-\beta E_j}}{\sum_{j=1}^{n} e^{-\beta E_j}}
\label{eq:prob-thermo}
\end{equation}
where $x_j^* = N_j^*/N$ and $N$ is the total number of molecules (the * denotes the equilibrium outcome). The partition function $Z$ given by
\begin{equation}
Z = \sum_{j=1}^{n} e^{-\beta E_j}
\label{eq:partition-thermo}
\end{equation}
and $x_j^*$ can also be written as
\begin{equation}
x_j^* = \frac{1}{Z} e^{-\beta E_j}
\label{eq:prob-thermo-Z}
\end{equation}

Readers familiar with statistical mechanics would immediately recognize that the second term in Eq.~\ref{eq:potential-thermo} corresponds to entropy ($S$)~\cite{venkat2017book}. Therefore, as observed by Venkatasubramanian~\cite{venkat2017book,venkat2025jaynes}, we recognize that by maximizing $\phi$, we are indeed maximizing the entropy $S$ subject to the \textit{energy constraint} given by the first term in Eq.~\ref{eq:potential-thermo}. \\

In general, in statistical teleodynamics, the effective utility of a goal-driven agent is a benefit-cost trade-off function of the form given by~\cite{Venkat2025activematter, venkat2025social}
\begin{equation}
    h_{j} = u_{j} - v_{j} - \ln N_j
    \label{eq:utility-generic}
\end{equation}

where $u_{j}$ is the benefit, $v_{j}$ is the cost, and $-\ln N_j$ is the disutility of competition among the agents. Table \ref{tab:utility-domains} lists some examples from different domains~\cite{venkat2025jaynes, venkat2025social} ($\beta$ represents the Boltzmann factor in thermodynamics, and the other Greek parameters represent different domain-specific concepts for different applications. Similarly, the variables $r_j, l_j, H$ and $S_j$ represent domain-specific concepts that are not relevant for our present discussion). \\

The generic teleodynamic potential is given by
\begin{eqnarray}
    \tilde{\phi} (\boldsymbol{x}) = \sum_{j=1}^n \int\left[ u_{j} - v_{j} - \ln x_j \right] d x_j 
    \label{eq:potential-generic}
\end{eqnarray}
The generic teleodynamic partition function is given by
\begin{equation}
\tilde{Z} =  \sum_{j = 1}^{n} \exp (u_j - v_j)
\label{eq:partition-generic}
\end{equation}
and the generic teleodynamic probability distribution is
\begin{equation}
x_j^* = \frac{1}{\tilde{Z}} \exp (u_j - v_j)
\label{eq:prob-generic}
\end{equation}

As shown in Eq.~\ref{eq:potential-thermo}, the $- \ln N_j$ term leads to entropy $S$ upon integration, and the $u_j$ and $v_j$ terms (upon integration) correspond to the \textit{constraints} that maximum entropy must satisfy. Thus, the persistence preferences of goal-driven agents, encoded in the terms $u_j$ and $v_j$, manifest themselves as constraints on maximum entropy. From Table~\ref{tab:utility-domains}, for example, we have (i) for gas molecules, $u_j = 0$ and $v_j = \beta E_j$, (ii) for income distribution in economics, $u_j = \alpha \ln S_j$ and $v_j = \eta (\ln S_j)^2$, and (iii) for segregation dynamics in sociology, $u_j = \alpha N_j + \ln(H-N_j)$, $v_j = \eta N_j^2$. These different mathematical forms of constraints (i.e., $\sum_{j=1}^n \int\left[ u_{j} - v_{j} \right] d x_j$) for maximum entropy across various domains show how different persistence effects can appear mathematically in our framework. Eq.~\ref{eq:partition-generic} and \ref{eq:prob-generic} could get complicated if $u_j$ and $v_j$ depend on $N_j$, as in the case of bird flocking and social segregation games shown in Table~\ref{tab:utility-domains}. However, this is not the case for the Thermodynamic and Income Games. As a result, we can obtain the probability distributions analytically, cleanly - e.g., exponential (energy) and lognormal (income) distributions. 

\begin{table}[!h]
    \centering
    \caption{Effective utility in different domains}
    \label{tab:utility-domains}
    \scalebox{0.9}
    {\begin{tabular}{c c c }
    \hline
        \textbf{Domain}& \textbf{System}& \textbf{Effective Utility} ($h_j$) \\ \hline\\
         Physics & Thermodynamics & $-\beta  E_j - \ln N_j$ \\\\
         Physics & Janus Particles & $-\dfrac{\omega r_j^a}{a} - \ln N_j$ \\\\
         Ecology & Ant crater formation & $b - \dfrac{\omega r_j^a}{a} - \ln N_j$ \\\\
         Ecology & Birds flocking & $\alpha N_j - \eta N_j^2+ \gamma N_j l_j - \ln N_j$ \\\\
         Sociology & Segregation dynamics & $\alpha N_j - \eta N_j^2 + \ln(H-N_j) -\ln N_j$ \\\\        
         Economics & Income distribution & $\alpha \ln S_j - \eta \left(\ln S_j\right)^2 - \ln N_j$ \\\\ 
         \hline
    \end{tabular}}
\end{table}

\section{Persistence and Memory in Cosmology}

Building on our statistical teleodynamic formulation of persistence and memory, in this section, we define them in a cosmological context. Although the Big Bang is not like an explosion, the following example is nonetheless intuitively helpful. Consider a familiar, if disturbing, account of a bomb going off in the middle of a large crowd of people. As soon as the bomb explodes, people start scattering in different directions. Several minutes after the explosion, people are still trying to escape in whatever way they can. Some are running, some are jumping into a car, some are riding a bike, some are trying to hide in a building, and so on. To remain safe, people do different things depending on what their local constraints allow. If the destruction is massive, people may try to escape to safety even hours or days after the explosion. Thus, days, months, or even years after a catastrophic event, people remember the event and its consequences. The event creates a \textit{generic} systemic memory that persists over time. It also creates \textit{specific} memories for different people of how they reacted, how they escaped, and how it forever changed their lives. They all share the same systemic memory of the explosion, but they all remember the specific details differently based on their individual experiences. The point of this example is that catastrophic events leave both generic and specific signatures on the elements of a system that persist for a long time. \\

Similarly, we suggest that the Big Bang and subsequent inflationary events (BBI) are of such cosmic magnitude, both literally and figuratively, that they left a persistent memory in the dynamics and evolution of the universe. The sudden inflationary expansion by about $10^{30}$ in scale, which lasted only from $10^{-36}$ to $10^{-32}$ seconds, was such an extraordinary cosmic shock that the universe still "remembers" it. The intrinsic quantum fluctuations that occurred during BBI produced an uneven spacetime structure. These non-uniformities persisted and became enormously magnified during inflation and the subsequent expansion of the universe. These later evolved into galactic proportions seen in the current macroscopic structure of the universe. Our main insight is that the BBI not only left its signature in the Cosmic Microwave Background (CMB), but also left a similar, indelible, uneven structures on the very fabric of the cosmos, namely, the spacetime. \\

We propose that the BBI left two kinds of persistent structural signatures on the spacetime fabric, one generic and one specific. Every element of spacetime remembers the BBI, but each in its own way. The generic signature is like a background effect that influences everything in the same manner, while the specific signature allows stellar objects to evolve according to their local histories and constraints. This is what we model as the teleodynamic bias potential $\Phi(t, x)$ in the next section, which we decompose into its generic background part ($\bar{\Phi}(t)$) and the specific fluctuations part ($\varphi(t, x)$), given by
\begin{equation}
\Phi(t, x) = \bar{\Phi}(t) + \varphi(t, x)
\label{eq:bias-background-local}
\end{equation}

In the teleodynamic framework, memory ($\Phi(t, x)$) is not a psychological or informational construct, but a physical property of an evolving cosmic system. It represents the persistent imprint left on the large-scale gravitational field by the BBI, the history of structure formation, matter flows, and tidal interactions across cosmic timescales. Because gravity in an expanding universe is long-range and non-ergodic, the distribution of matter does not ``forget” its past configurations, including the BBI, which is to say that we cannot treat the physics here to be completely Markovian. Instead, every merger, filamentary collapse, void expansion, or tidal alignment leaves traces in the statistical and systemic organization of the cosmic web, and these persistent features accumulate into a global bias functional, which we refer to as cosmic memory. \\

From a dynamical perspective, cosmic memory encodes how the universe departs from the assumptions of perfect equilibrium that underlie standard statistical mechanics. This is exemplified by the fact that a collisionless, self-gravitating system never fully thermalizes, and its phase-space distribution maintains long-lived correlations that cannot be erased by expansion alone. These correlations are precisely what teleodynamics captures. As the universe evolves, preferred organizational patterns survive Hubble timescales and feed back into both the background expansion and the growth of structure. In this sense, we say that memory arises as an emergent, non-local statistical field that quantifies how past gravitational organization biases affect evolution. \\

Such process-dependent memory effects on macroscopic structure and properties are not unique to cosmology. The formation and evolution of macroscopic structures in materials science have similar features. For example, in the design, formulation, and manufacturing of rubber-based engineered materials, small microscopic irregularities, such as fluctuations in temperature and stress profiles during the curing process, can transform into large macroscopic variations in structure, properties, and behavior. This micro-to-macro mathematical modeling is complicated by interactions between physical and chemical transformations arising from complex sulfur-chemistry reactions, as discussed by Ghosh et al.~\cite{Ghosh2003rubber}. Of course, one needs to be careful, as all analogies have limitations. However, structural formation and evolution in material science can provide valuable insight into what to expect in other domains. 
\\
\\
Cosmologically speaking, this memory becomes especially important once structure formation becomes nonlinear, as is the case in the current age of the universe. Filaments, clusters, and voids act as reservoirs of persistent tidal information, storing the integrated effects of billions of years of gravitational interactions. This accumulated organization cannot be described by local density alone; instead, it manifests itself as a global functional of the cosmic web. Viewed this way, cosmic memory is not an additional field or exotic substance introduced by hand, but is instead the intrinsic statistical and systemic outcome of a universe that has evolved under long-range gravity, formed a hierarchy of structures, and retained correlations across cosmic timescales. Teleodynamics formalizes this insight into a predictive framework: the accumulated memory of structure formation influences large-scale dynamics, shapes late-time acceleration, and modulates the growth of perturbations. In this sense, cosmic memory is the physical bridge connecting the microscopic history of gravitational organization and the macroscopic observables that define modern cosmology.

\section{Statistical Teleodynamics of Cosmology}

Guided by these considerations and the mathematical models of persistence and memory in other domains, as shown in Table~\ref{tab:utility-domains}, we are now ready to develop a teleodynamic framework for cosmology that captures the effects of persistent structures, memory effects, and dynamically induced preferences. We start with Eq.~\ref{eq:utility-generic}, Eq.~\ref{eq:partition-generic}, and Eq.~\ref{eq:prob-generic}, considering a discrete space of microstates $x$. Instead of assigning weights solely by the energy $E(x)$, as in Eq.~\ref{eq:utility-thermo}, we introduce an additional functional $\Phi(x)$ which captures a path-dependent or environment-dependent persistence bias as a constraint intrinsic to the system. Therefore, we have $u(x) = 0$ and $v(x) = \beta E(x) + \alpha \Phi(x)$, yielding
\begin{equation}
h(x)= -\beta E(x) - \alpha \Phi(x) - \ln N(x)
\label{eq:utility-cosmo}
\end{equation}

The microscopic probability distribution is given by 
\begin{equation}
p(x) = \frac{1}{Z} \exp[-\beta E(x) - \alpha \Phi(x)]
\label{eq:prob-cosmo-persistence}
\end{equation}
with the partition function
\begin{equation}
Z(\beta, \alpha) = \sum_x \exp[-\beta E(x) - \alpha \Phi(x)]
\label{eq:partition-cosmo-persistence}
\end{equation}

The parameter $\alpha$ controls the strength of the teleodynamic bias, while the functional $\Phi$ can depend on nonlocal correlations, environmental constraints, dynamical stability conditions, and statistical preferences linked to the history of the system. In the standard thermodynamic limit where $\Phi$ vanishes, one recovers the ordinary canonical ensemble. \\

Since it is natural to use free energy instead of the game-theoretic potential in cosmology, we will now switch to a free energy-based analysis. As Venkatasubramanian showed~\cite{venkat2017book, Venkat2025activematter, venkat2025jaynes}, minimizing free energy is equivalent to maximizing game-theoretic potential $\phi$. \\

To make the structure of the teleodynamic free energy ($F_{\rm TD}$) explicit, it is useful to derive it directly from the microscopic probability distribution. We begin with the expression \eqref{eq:prob-cosmo-persistence} which implies
\begin{equation}
\ln p(x)=-\ln Z-\beta E(x)-\alpha \Phi(x)
\end{equation}
Multiplying this expression by $p(x)$ and summing over all microstates gives
\begin{equation}
\sum_x p(x)\ln p(x)=\sum_x p(x)\big[-\ln Z-\beta E(x)-\alpha \Phi(x)\big]
\end{equation}
Using the normalization condition $\sum_x p(x)=1$, this becomes
\begin{equation}
\sum_x p(x)\ln p(x)=-\ln Z - \beta \sum_x p(x)E(x) - \alpha \sum_x p(x)\Phi(x)
\end{equation}
Solving this relation for $\ln Z$ yields
\begin{equation}
\ln Z = - \sum_x p(x)\ln p(x) - \beta \sum_x p(x)E(x) - \alpha \sum_x p(x)\Phi(x)
\end{equation}
Inserting this into the definition of the teleodynamic free energy $F_{\rm TD} = -\frac{1}{\beta}\ln Z$, gives us the following
\begin{equation}
F_{\rm TD} = \frac{1}{\beta}\sum_x p(x)\ln p(x) + \sum_x p(x)E(x) + \frac{\alpha}{\beta}\sum_x p(x)\Phi(x)
\end{equation}
This expression makes clear that the equilibrium distribution minimizes a combination of energetic, entropic and teleodynamic contributions, with the final term encoding the effects of persistent memory and structural bias that have no analogue in ordinary thermodynamics.
 The third contribution has no analog in ordinary thermodynamics and encodes the persistence of organizational constraints and dynamical feedback not reducible to energy alone.\\

The evolution equations of teleodynamic systems follow by considering the probability flux in the configuration space. So, if $x(t)$ denotes microscopic variables, the coarse-grained dynamics acquires an additional drift proportional to the teleodynamic potential. In a gradient flow approximation, the equation of motion takes the form
\begin{equation}
\dot{x} = -\nabla E(x) - \alpha \nabla \Phi(x) + \eta(t)
\end{equation}
where $\eta$ represents stochastic fluctuations and the deterministic term $-\alpha\nabla\Phi$ drives the system toward attractors defined not only by the minima of the energy but also by the minima of the teleodynamic functional. This allows us to naturally generate stable macroscopic tendencies that would require fine-tuning in conventional thermodynamics, and a probability-based formulation gives us a teleodynamic Fokker-Planck equation
\begin{equation}
\frac{\partial p(x, t)}{\partial t} = \nabla \cdot \left[ p(x, t) \nabla\left(E(x) + \alpha\Phi(x)\right) \right] + D\nabla^2 p(x, t)
\end{equation}
demonstrating again that teleodynamics introduces a functional bias, altering both the drift and the stationary solutions. \\

The structure of this whole theory consists of a few vital ingredients: first is a generalized Gibbs distribution that retains the form of an exponential family but includes an additional functional in the exponent. The second is the appearance of an effective potential $E_{\text{eff}}(x) = E(x) + \alpha\Phi(x)$ that modifies the free energy landscape and the deterministic drift of coarse-grained dynamics. The third is the recognition that complex and evolving macroscopic systems may not be described by equilibrium thermodynamics, but rather by a combination of free-energy minimization and attractor selection, governed by $\Phi$. These features become particularly important in cosmology, where long-range gravitational forces, nonlocality, and the presence of large-scale correlated structures imply that the assumptions of ordinary statistical mechanics are not satisfied. In such a context, the teleodynamic corrections provide a natural mechanism for introducing effective energy densities, biasing large-scale flows, and generating attractor behavior without invoking new particle species or additional fundamental fields. The key differences between thermodynamics and teleodynamics are summarized in Table~\ref{tab:differences-thermo-teleo} and in Eq.~\ref{eq:potential-generic}-\ref{eq:prob-generic}. 

\begin{table}[h]
\centering
\begin{tabular}{|l|l|l|}
\hline
Property & Ordinary thermodynamics & Teleodynamics \\
\hline
Microscopic weight & $e^{-\beta E(x)}$ & $e^{-\beta E(x)-\alpha\Phi(x)}$ \\
Equilibrium criterion & minimize free energy & minimize $F + \alpha\langle\Phi\rangle$ \\
Assumptions & ergodic, memoryless, short range & non-ergodic, memory-influenced, nonlocal \\
Attractors & energy minima only & minima of $E + \alpha\Phi$ \\
Dynamics & $\dot{x} = -\nabla E + \eta$ & $\dot{x} = -\nabla(E + \alpha\Phi) + \eta$ \\
\hline
\end{tabular}
\caption{Key differences between thermodynamics and teleodynamics}
\label{tab:differences-thermo-teleo}
\end{table}

\section{Cosmological Teleodynamics Framework}

The teleodynamic description of cosmological dynamics begins with the recognition that the evolution of a macroscopic system is determined not only by instantaneous energies but also by the statistical weight of its entire history. So we start here by considering the appropriate statistical object to be the maximum--caliber distribution on histories $\Gamma = \{x(t), p(t)\}$ which is defined by
\begin{equation} \label{maxcab}
\mathcal{P}[\Gamma] \propto \exp[-\beta \mathcal{A}[\Gamma] - \alpha \mathcal{K}[\Gamma]]
\end{equation}
where $\mathcal{A}[\Gamma]$ is the action functional that captures the usual dynamical contributions and $\mathcal{K}[\Gamma]$ is a bias functional that encodes the memory, environmental dependence, and structural correlations that the system retains over time. The parameters $\beta$ and $\alpha$ enforce the macroscopic constraints on energy and teleodynamic bias, which are kind of like Lagrange multipliers of maximum entropy. To make this explicit, we write the mechanical part of the history weight as
\begin{equation}
\mathcal{A}[\Gamma] = \int dt \left[ \frac{p^2}{2a^2m} + m \Psi(x, t) \right]
\end{equation}
where $\Psi$ is the gravitational potential, and the teleodynamic part as
\begin{equation}
\mathcal{K}[\Gamma] = \int dt\, \Phi\left(x(t), p(t); f, \mathcal{E}\right)
\end{equation}
where $\Phi$ is the bias functional depending on the configuration variables, the phase-space distribution $f$, and the coarse environmental fields $\mathcal{E}$ such as the tidal tensor or the surrounding density field. The functional $\Phi$ here accounts for nonlocal correlations, dynamical memory, and persistent environmental structure, and is the core ingredient that distinguishes teleodynamics from ordinary thermodynamics. \\

The effective equations of motion are derived by extremizing the exponent of $\mathcal{P}[\Gamma]$, and the variational differentiation with respect to $x(t)$ and $p(t)$ yields the drift equations
\begin{equation}
\dot{x} = \frac{p}{a^2m}
\end{equation}
\begin{equation}
\dot{p} = -\nabla\Psi(x, t) - \alpha \nabla\Phi(x, p, t) - Hp
\end{equation}

This is helpful as it shows that the dynamics experiences an additional force proportional to the gradient of the bias functional. This force is not sourced from new matter fields but rather represents a coarse-grained structural response of the system to its own past evolution and environment. As a result, the presence of $\alpha\nabla\Phi$ alters the gravitational streaming of matter, modifies the effective potential, and influences the background expansion. By promoting the trajectory equations to a kinetic description, one obtains the teleodynamic Boltzmann equation, from which the modified continuity, Euler, and growth equations follow. \\

In a cosmological setting, the bias functional $\Phi(t,x)$ enters the maximum–caliber weight through the history action as an additional structural term influencing the statistical evolution of cosmic phase space. Since the universe on large scales is well described by an FLRW background with small fluctuations about homogeneity and isotropy, it is natural to separate the functional into its spatially averaged component and its inhomogeneous deviations, and so we can write
\begin{equation}
\Phi(t,x)=\bar{\Phi}(t)+\varphi(t,x)
\tag{\ref{eq:bias-background-local}}
\end{equation}
It is important to understand how and why this decomposition works. It follows the same logic that underlies the treatment of any scalar field or effective quantity in cosmology, wherein the spatial average of the field governs the background expansion, while its fluctuations source clustering and local gravitational potentials. Because the teleodynamic bias functional can depend nonlocally on the matter distribution, the tidal environment, or the accumulated memory of the cosmic web, it can vary both temporally and spatially. One notes that the FLRW symmetries imply that only the spatial average of such a quantity can enter the background equations, while its space-dependent part must enter the perturbed sector. So, the decomposition (\ref{eq:bias-background-local}) is not an assumption, but a direct consequence of applying the standard cosmological splitting to a quantity that affects both the background and perturbations.\\

Let us consider this split clearly, beginning with the Einstein–Hilbert action for gravity coupled to matter. We include the teleodynamic bias as an additive functional in the effective action obtained from coarse-graining over microscopic histories, and the result of this can be written schematically as
\begin{equation}
S_{\rm eff}[g_{\mu\nu},\Psi_m] = \int d^4x \sqrt{-g}\left[\frac{M_P^2}{2}R + \mathcal{L}_m(g_{\mu\nu},\Psi_m)\right] - \alpha \int d^4x \sqrt{-g}\,\Phi[g_{\mu\nu},\Psi_m;\mathcal{E}]
\label{eq:effective-action}
\end{equation}
where $g_{\mu\nu}$ is the spacetime metric, $R$ is the Ricci scalar, $M_P$ is the reduced Planck mass, $\Psi_m$ collectively denotes the degrees of freedom of matter and $\mathcal{L}_m$ is the Lagrangian matter density, and the parameter $\alpha$ controls the strength of the teleodynamic contribution. The last term arises from the coarse-grained limit of the maximum–caliber path weight and represents the macroscopic imprint of nonlocal memory on the spacetime dynamics. $\Phi$ is the teleodynamic bias functional, which encodes a coarse-grained imprint of persistence, memory, and environment dependence, and the symbol $\mathcal{E}$ denotes collective environmental variables or invariants that can be constructed from the distribution of matter. We can now vary this action with respect to $g_{\mu\nu}$, which gives us modified Einstein equations
\begin{equation}
M_{\rm P}^2 G_{\mu\nu}=T_{\mu\nu}^{\rm m}+T_{\mu\nu}^{\rm TD}
\end{equation}
where the teleodynamic stress tensor is given as
\begin{equation}
T_{\mu\nu}^{\rm TD} = -\frac{2}{\sqrt{-g}}\,\frac{\delta}{\delta g^{\mu\nu}}\Biggl(\alpha\int d^4x \sqrt{-g}\,\Phi(t,x)\Biggr)
= \alpha\,\Phi\, g_{\mu\nu} - 2\alpha\,\frac{\delta \Phi}{\delta g^{\mu\nu}}
\label{eq:teleo-stress-general}
\end{equation}
We note here that the first term in \eqref{eq:teleo-stress-general} behaves as an effective perfect-fluid contribution, while the second term encodes the possible metric dependence of the bias functional associated with tidal correlations or geometric memory. If $\Phi$ depends only on cosmic time at the background level, then $\delta \Phi/\delta g^{\mu\nu}=0$ for the homogeneous mode, and we obtain a perfect fluid form, which is
\begin{equation}
T^{\mu}{}_{\nu\,{\rm TD}} = {\rm diag}[-\rho_{\rm TD},\,p_{\rm TD},\,p_{\rm TD},\,p_{\rm TD}]
\end{equation}
with
\begin{equation}
\rho_{\rm TD}(t) = \alpha\,\bar{\Phi}(t), \qquad p_{\rm TD}(t) = \alpha\,\bar{\Pi}_{\rm TD}(t)
\label{eq:teleo-background-rho}
\end{equation}
where $\bar{\Pi}_{\rm TD}(t)$ arises from the time dependence of $\bar{\Phi}(t)$ when inserted into the covariant conservation equation $\nabla_\mu T^{\mu\nu}_{\rm TD}=0$. These expressions show that the homogeneous component $\bar{\Phi}(t)$ behaves as an additional energy density and pressure at the background level. This can eventually modify the Friedmann equations while respecting the cosmological symmetries, as we shall see soon. This establishes clearly why a spatially averaged bias functional must influence the expansion history, as only its homogeneous mode survives in the background of an FLRW universe.\\

Now we would like to consider the inhomogeneous part $\varphi(t,x)$, which affects the perturbed sector and therefore determines the local clustering behavior. In the Newtonian gauge, the perturbed Einstein equations imply that the Poisson equation for the potential $\Psi$ becomes
\begin{equation}
\nabla^2\Psi = 4\pi G a^2 \bigl(\bar{\rho}\,\delta + \delta\rho_{\rm TD}\bigr)
\end{equation}
with
\begin{equation}
\delta\rho_{\rm TD}(t,x) = \alpha\,\varphi(t,x) + \text{metric response terms}
\end{equation}
When the metric dependence of the bias functional is negligible on subhorizon scales, the dominant contribution can be identified as being
\begin{equation}
\delta\rho_{\rm TD}(t,x) \simeq \alpha\,\varphi(t,x)
\end{equation}
This will act as an emergent clustering density, as we shall see later. In Fourier space, this becomes
\begin{equation}
\delta\rho_{\rm TD}(k,t)=\frac{\alpha}{4\pi G a^2}\,k^2 \varphi_k(t)
\end{equation}
so that the effective potential satisfies
\begin{equation}
\nabla^2\Psi_{\rm eff} = \nabla^2\bigl(\Psi + \alpha\,\varphi\bigr)
= 4\pi G a^2 \bigl(\bar{\rho}\,\delta + \delta\rho_{\rm TD}\bigr)
\end{equation}
This shows that the inhomogeneous component of the bias functional provides an additional source of gravitational potential, modifying galaxy motions, etc., thereby demonstrating that the homogeneous and inhomogeneous parts of $\Phi$ play distinct physical roles. While $\bar{\Phi}(t)$ governs the background expansion through its contribution to the stress tensor, $\varphi(t,x)$ shapes the spatial variations in the effective potential and thus modifies clustering. This division is unavoidable once one acknowledges that memory, environmental dependence, and persistent structure necessarily generate both large-scale coherent contributions and spatially varying corrections. The decomposition (\ref{eq:bias-background-local}) is therefore a natural and physically required feature of any teleodynamic description of cosmology.Please note that extended derivations, clarifications, and a deeper discussion on the fundamental aspects of the teleodynamic action, the field decomposition, its perturbations, and gauge invariant properties are in the appendix. The appendix also clearly establishes how Cosmological Teleodynamics produces a cosmology which is fundamentally distinct from a simple barotropic fluid cosmology. \\

Also note that in the teleodynamic interpretation developed in the following sections, the "agents" of the cosmic game are not individual particles, stars, or local structures, but galaxies understood as coarse-grained units that propagate long-lived dynamical memory. This choice follows naturally from the fact that galaxies, rather than microscopic constituents, are the fundamental carriers of persistent organization in the late Universe as they preserve tidal histories, respond coherently to their large-scale environments, and act as the basic building blocks whose collective interactions generate the cosmic web. Treating galaxies as agents, therefore, reflects their role as the minimal scale at which non-ergodicity, correlation persistence, and teleodynamic bias manifest in a physically meaningful way. Thus, galaxies belong to the third category of agents, defined in statistical teleodynamics as intrinsically persistent systems whose behavior arises from physical constraints, such as long-range gravitational memory and accumulated structure, rather than from intention or cognition. Therefore, galaxies participate in a potential game-like evolution, with their aggregate dynamics shaping the emergent teleodynamic forces that drive various aspects of cosmology, as we shall see below.

\subsection{Boltzmann Dynamics and Galactic Phase-space Evolution}

The starting point for any analysis of matter evolution in cosmology is the Boltzmann equation of collisionless tracers. The Boltzmann equation considers the evolution of matter for the phase-space distribution $f(t,\mathbf{x},\mathbf{p})$ where $\mathbf{p}=a m \mathbf{v}$ denotes the canonical momentum of a collisionless tracer of mass $m$. In comoving coordinates under the assumption that gravity is the only long–range force and that the system possesses no dynamical memory,  the Liouville flow allows one to write the standard collisionless Boltzmann equation as \cite{bol1dodelson2020modern,bol2Enomoto:2023cun,bol3Bazow:2016oky,bol4lee2016asymptotic,bol5cercignani2002relativistic,bol6coble1997dynamical,bol7weinberg2008cosmology,bol8shoji2010massive}
\begin{equation}
\frac{\partial f}{\partial t}
+ \frac{\mathbf{p}}{a^{2} m}\cdot\nabla_{\mathbf{x}} f
- \nabla_{\mathbf{x}}\Psi \cdot \nabla_{\mathbf{p}} f
- H\, \mathbf{p}\cdot\nabla_{\mathbf{p}} f
= 0
\end{equation}
where $\Psi$ is the Newtonian potential determined from the Poisson equation. This form implicitly assumes ergodicity, local relaxation, and the absence of long–lived correlations. However, cosmological structure formation violates all of these assumptions, as matter resides in a cosmic web whose tidal fields and filamentary geometry introduce persistent correlations and memory at all scales. Teleodynamics accommodates this by introducing a nonlocal, memory–carrying drift term. The microscopic bias functional is decomposed as
\begin{equation*}
\Phi(t,\mathbf{x}) = \bar{\Phi}(t) + \varphi(t,\mathbf{x})
\tag{\ref{eq:bias-background-local}}
\end{equation*}
where $\varphi$ encodes spatially varying correlations and environmental response. The total acceleration then becomes
\begin{equation}
\mathbf{a} = -\nabla\Psi - \alpha\,\nabla\Phi
\end{equation}
so that the teleodynamic Boltzmann equation is
\begin{equation}
\frac{\partial f}{\partial t}
+ \frac{\mathbf{p}}{a^{2} m}\cdot\nabla_{\mathbf{x}} f
- \left(\nabla_{\mathbf{x}}\Psi + \alpha\nabla_{\mathbf{x}}\Phi\right)\cdot \nabla_{\mathbf{p}} f
- H\,\mathbf{p}\cdot\nabla_{\mathbf{p}} f
= 0
\end{equation}
The term proportional to $\alpha$ represents the coarse-grained effect of long-lived correlations, environmental structure, and nonlocal memory encoded by the bias functional. Its presence here reshapes the flow in the phase space and generates an effective potential $\Psi_{\text{eff}} = \Psi + \alpha\Phi$ that governs the trajectories of matter elements. To understand the implications for galactic dynamics, we first compute the velocity moments of this kinetic equation. The continuity equation remains
\begin{equation}
\dot{\delta} + \frac{1}{a}\nabla \cdot \left[(1 + \delta)u\right] = 0
\end{equation}
but the Euler equation changes to
\begin{equation}
\dot{u} + Hu + \frac{1}{a}(u \cdot \nabla)u = -\frac{1}{a}\nabla\Psi - \frac{1}{a(1 + \delta)}\nabla \cdot P - \frac{\alpha}{a}\nabla\Phi
\end{equation}
where $P$ denotes our stress tensor. Linearizing the continuity and Euler equations and transforming to Fourier space leads to the coupled system
\begin{equation}
\dot{\delta}(k,t) = -\frac{1}{a}\,\theta(k,t),
\end{equation}
\begin{equation}
\dot{\theta}(k,t) + H\theta(k,t) - 4\pi G a \bar{\rho}\,\delta(k,t)
= -\frac{c_{s}^{2} k^{2}}{a}\,\delta(k,t)
- \frac{\alpha k^{2}}{a}\,\Phi_{k}(t)
\end{equation}
showing explicitly that the teleodynamic contribution provides a new scale–dependent source term for the velocity divergence. This term may behave effectively as a renormalization of the gravitational coupling or as a dispersive pressure-like correction depending on the structure of the kernel relating $\Phi_{k}$ to the matter distribution. Depending on the structure of the response $\Phi_k$, it is clear that the growth of perturbations acquires either an effective renormalization of the gravitational coupling or an additional pressure-like dispersive term. These corrections propagate directly to the peculiar velocity statistics \cite{pec1peebles1976peculiar,pec2graziani2019peculiar}, the velocity-density cross-spectrum, and the bulk flow variance, and in particular, the relation between the divergence of the velocity field and the matter overdensity becomes
\begin{equation}
\theta(k, a) = -aH f(k, a) \delta(k, a)
\end{equation}

This is a scale-dependent growth rate, $f(k, a)$, that departs from the scale independence predicted by the standard collisionless cold dark matter picture. \\

We see that the teleodynamic modification of the Boltzmann equation, therefore, induces characteristic signatures in galaxy motions and peculiar velocities. The additional drift term produces velocity fields with enhanced coherence over large distances, non-Gaussian velocity distributions, and a partial decorrelation between velocity and density, all of which differ from the predictions of an idealized gravitational gas governed solely by $\Psi$. Observations of bulk flows, pairwise velocities, and redshift-space distortions already hint at such anomalies, suggesting that a framework incorporating nonlocal memory and structural bias may be better suited to describing galactic phase-space dynamics. Unlike the ideal gas approximation, which assumes that particles respond only to instantaneous gravitational forces, the teleodynamic picture accounts for the fact that galaxies evolve within a network of long-range correlations created by the cosmic web and retain dynamical memory of their environment. This makes the teleodynamic Boltzmann framework a more realistic representation of large-scale cosmic motion and establishes the foundation for the remaining ingredients of teleodynamic cosmology.

\subsection{Friedmann Dynamics and Dark Energy}
The Friedmann equations, as we know, describe the dynamics of an expanding universe in a homogeneous and isotropic FLRW spacetime. In the standard picture, one assumes that the background energy density is given solely by physical matter components so that the expansion obeys
\begin{equation}
3M_{P}^{2}H^{2} = \rho,
\label{eq:StandardF1}
\end{equation}
\begin{equation}
-2M_{P}^{2}\dot{H} = \rho + p
\label{eq:StandardF2}
\end{equation}
Together with the continuity equation
\begin{equation}
\dot{\rho} + 3H(\rho + p) = 0,
\label{eq:StdContinuity}
\end{equation}
these relations fully determine the FLRW background given a set of matter components with equations of state. One immediate consequence of \eqref{eq:StandardF1}-\eqref{eq:StandardF2} is that cosmic acceleration 
\begin{equation}
\frac{\ddot{a}}{a} > 0,
\label{eq:accelcondinformal}
\end{equation}
thus requires a dominant contribution with sufficiently negative pressure. In conventional $\Lambda$CDM, this condition is met by postulating a cosmological constant or suitably tuned scalar fields with an equation of state $w \simeq -1$. Teleodynamics provides an alternative interpretation of acceleration by identifying the missing component not as new fundamental fields, which has become the prevailing trend in cosmology, but instead as a statistically emergent effect of dynamical memory in a self-gravitating medium. To make this precise, recall that the teleodynamic formulation assigns to every microscopic history a weight of the form \eqref{maxcab} where the bias functional captures nonlocal correlations, memory, and the influence of long-lived structures. At the coarse-grained level, this functional can then split naturally into a homogeneous and an inhomogeneous piece by writing
\begin{equation*}
\Phi(t,\mathbf{x}) = \bar{\Phi}(t) + \varphi(t,\mathbf{x})
\tag{\ref{eq:bias-background-local}}
\end{equation*}
with a vanishing spatial mean of the inhomogeneous part and the average $\bar{\Phi}(t)$ enters the background action through an effective contribution to the coarse-grained stress tensor. This leads to a modification of the Friedmann sector to
\begin{equation}\label{eq:TDF1}
3M_P^2 H^2 = \rho + \alpha \bar{\Phi}(t)
\end{equation}
and
\begin{equation} \label{eq:TDF2}
-2M_P^2 \dot{H} = \rho + p + \alpha \Pi_{\text{TD}}(t)
\end{equation}
where $\Pi_{\text{TD}}(t)$ signifies the effective pressure associated with the temporal evolution of $\bar{\Phi}$. The structure of these equations shows that the teleodynamic bias contributes an additional homogeneous component to the cosmic energy budget, with the effective density and pressure given by
\begin{equation}\label{eq:rhoPTD}
\rho_{\text{TD}}(t) = \alpha \bar{\Phi}(t), \quad p_{\text{TD}}(t) = \alpha \Pi_{\text{TD}}(t)
\end{equation}
Note that no additional fields are required for this term to appear. It follows directly from the statistical and systemic coarse-graining of matter in a gravitationally correlated environment and represents the macroscopic imprint of dynamical memory and nonlocal structure. One can derive \eqref{eq:TDF1} and \eqref{eq:TDF2} directly by inserting the modified stress tensor
$$T^{\mu\nu}_{\mathrm{eff}} = T^{\mu\nu}_{\mathrm{matter}} + T^{\mu\nu}_{\mathrm{TD}}$$
with $T^{\mu\nu}_{\mathrm{TD}}$ determined through variational differentiation of the coarse-grained teleodynamic functional
\begin{equation}
    S_{\mathrm{eff}}[g_{\mu\nu}] = \int d^{4}x\,\sqrt{-g}\left[\mathcal{L}_{\mathrm{matter}} - \alpha\,\Phi\right]
\end{equation}
Spatial averaging then ensures that only $\bar{\Phi}(t)$ contributes to the background energy density and only $\Pi_{\mathrm{TD}}(t)$ contributes to the effective pressure. \\

The condition for accelerated expansion is obtained by combining the Friedmann equations in the usual way, as
\begin{equation}
\frac{\ddot{a}}{a} = -\frac{1}{6M_P^2}[\rho + 3p + \rho_{\text{TD}} + 3p_{\text{TD}}] > 0
\end{equation}

In the absence of teleodynamic corrections, this inequality demands an exotic fluid with $p \simeq -\rho$, which is the basis of the dark energy. But with teleodynamic contributions accounted for correctly, the accelerating condition may be satisfied even if the physical matter components obey normal equations of state. Furthermore, when the bias functional is sufficiently slowly varying, the effective equation of state satisfies
\begin{equation}
w_{\text{TD}} \equiv \frac{p_{\text{TD}}}{\rho_{\text{TD}}} = \frac{\Pi_{\text{TD}}(t)}{\bar{\Phi}(t)} \approx -1
\end{equation}
and the teleodynamic term behaves like a cosmological constant. More generally, the rate of change of $\bar{\Phi}$ can also determine whether the expansion is quintessence, phantom, or quintom-like. This shows that the entire late-time acceleration can be interpreted simply as a large-scale statistical and systemic effect of a memory-bearing bias functional in a self-gravitating universe. \\

Instead of introducing new scalar fields, potentials, or hidden matter sectors in our theory \cite{dem1Giare:2025pzu,dem2Marcy:2024dmo,dem3Wu:2025wyk}, one simply recognizes that the universe is not a system that relaxes to a gas-like equilibrium or one whose expansion is governed solely by local microphysics. The long-range nature of gravity and the persistent structures generated by the Big Bang and subsequent cosmic evolution imply that the background expansion should include contributions that have nonlocal correlations and path dependence. The teleodynamic energy density $\rho_{\text{TD}}$ arises precisely from these considerations and provides an emergent explanation for the cosmic acceleration. The term $\rho_{\mathrm{TD}} = \alpha\bar{\Phi}$ arises not from a new microscopic field, but from the coarse-grained effect of self-gravitating correlations and memory, which accumulate as the Universe forms persistent structures and departs more strongly from ergodicity. In this sense, we see that late-time acceleration emerges naturally as the Universe becomes increasingly non-thermal and the expansion is governed by an effective potential generated by the ensemble of long-lived gravitational correlations. Teleodynamics in this sense provides a statistical and systemic explanation for cosmic acceleration grounded in the physics of memory-bearing, nonlocally correlated matter rather than in additional dark sectors or fine-tuned vacuum energies.

\subsection{Dark Matter and Structure Growth}

In standard cosmology, the problem of dark matter is addressed by introducing a new, non-relativistic matter component whose density obeys the usual continuity and Euler equations and whose gravitational potential satisfies the Poisson equation. Teleodynamics now offers a different perspective in which the missing mass effect arises not from a new particle species or primordial remnants, but from the structural correction induced by the teleodynamic bias functional at the level of inhomogeneities. This emerges naturally once one decomposes the bias into its homogeneous and fluctuating parts again, as before
\begin{equation*}
\Phi(t,\mathbf{x}) = \bar{\Phi}(t) + \varphi(t,\mathbf{x})
\tag{\ref{eq:bias-background-local}}
\end{equation*}
and allows the inhomogeneous component to participate in the gravitational potential felt by matter elements.\\

The teleodynamic modification of the kinetic theory would imply that particles experience an effective potential consisting of the Newtonian part plus a correction generated by the bias functional. This potential is defined by
\begin{equation}
\Psi_{\text{eff}}(t, x) = \Psi(t, x) + \alpha \varphi(t, x)
\label{eq:psieff-psi-phi}
\end{equation}
with the gravitational potential obeying the usual Poisson equation
\begin{equation}
\nabla^2\Psi = 4\pi Ga^2\bar{\rho} \delta
\end{equation}

To determine the additional contribution, one applies the Laplacian to $\Psi_{\text{eff}}$, giving us
\begin{equation}
\nabla^2\Psi_{\text{eff}} = 4\pi Ga^2\bar{\rho} \delta + \alpha \nabla^2\varphi
\end{equation}

This motivates the definition of an emergent clustering density, which would be
\begin{equation}
\rho_{\text{TD}}(t, x) = \frac{\alpha}{4\pi Ga^2} \nabla^2\varphi(t, x)
\end{equation}
so that the effective potential satisfies the Poisson equation
\begin{equation}
\nabla^2\Psi_{\text{eff}} = 4\pi Ga^2[\bar{\rho} \delta + \rho_{\text{TD}}]
\label{eq:laplacian-psieff-rhoTD}
\end{equation}

Interestingly, the term $\rho_{\text{TD}}$ acts as a source of gravitational clustering, playing the role conventionally attributed to dark matter. Its origin is again statistical and systemic, rather than material.  It is derived from the curvature of the teleodynamic bias, which captures the influence of long-range correlations, tidal structure, or flow-dependent memory on the local gravitational environment. No new fields or particles are required for this term to emerge, and so it is an effective density derived from the nonlocal structure of the cosmic matter distribution.\\

The dynamical influence of $\rho_{\text{TD}}$ is made more explicit by examining the Euler equation for the peculiar velocity field derived from the teleodynamic Boltzmann equation, in which, after linearization in Fourier space, we can find that
\begin{equation}
\dot{\theta}(k, t) + H\theta(k, t) - 4\pi Ga \bar{\rho} \delta(k, t) = -\frac{\alpha k^2}{a}\varphi_k(t)
\end{equation}
which, upon using the definition of the effective density, becomes
\begin{equation}
\dot{\theta}(k, t) + H\theta(k, t) - 4\pi Ga [\bar{\rho} \delta(k, t) + \rho_{\text{TD}}(k, t)] = 0
\end{equation}

This shows that the teleodynamic term enhances the effective gravitational strength precisely as required to explain galactic rotation curves, cluster dynamics, and gravitational lensing. The enhancement is determined by the response kernel that relates $\varphi_k$ to matter fluctuations and their tidal environment, so the scale and time dependence of the effective clustering can be adapted to the observed phenomenology without invoking any particulate dark matter. The phenomenological versatility here becomes apparent once one specifies the form of the response kernel connecting $\varphi_k$ with the underlying matter perturbations, as in general we can write $\varphi_k = K(k, a) \delta_k$, with $K(k, a)$ signifying nonlocal environmental memory, tidal couplings, or velocity--space correlations. The effective density then takes the form $\rho_{\text{TD}}(k, a) = \Delta\mu(k, a) \bar{\rho} \delta(k, a)$, where $\Delta\mu(k, a) \equiv \alpha K(k, a)/(4\pi Ga^2)$. The scale and time dependence of $K$ determines the strength of clustering and thereby allows teleodynamic dark matter to mimic a wide range of popular dark matter phenomenologies. \\

For example, a kernel that tends to a constant on large scales but grows mildly toward smaller scales reproduces the behavior of cold dark matter, in which the gravitational coupling is nearly scale-independent across the observed linear regime. A kernel with a transition scale $k_{\text{fs}} \sim \ell_{\text{env}}^{-1}$, reflecting the characteristic environmental length $\ell_{\text{env}}$, generates suppressed clustering for $k > k_{\text{fs}}$ and thus imitates warm dark matter without invoking relics. If the kernel decreases for wavenumbers associated with dense environments, then $K$ induces a pressure-like contribution to the Euler equation, potentially yielding cored halo profiles reminiscent of self-interacting dark matter models. A kernel with an oscillatory or rapidly varying structure at small $k$ produces an effective scale-dependent modification to the force law that generates ultralight scalar dark matter effects.\\

The teleodynamic interpretation of dark matter is hence closely linked to the non-ergodic and nonlocal character of gravitational dynamics. In a universe where structure formation generates filaments, walls, and tidal fields extending over tens or hundreds of megaparsecs, the local gravitational force experienced by a galaxy cannot depend solely on the instantaneous Newtonian potential sourced by visible matter. So, the memory of the surrounding cosmic web, which is encoded in the bias functional, will contribute an additional drift proportional to $-\nabla\Phi$, which manifests itself as an effective mass distribution. The emergent density $\rho_{\text{TD}}$ will reflect this coarse-grained structural correction and provide a natural explanation for the observed gravitational anomalies. Unlike standard dark matter, the teleodynamic approach attributes the missing mass influence to the statistical system structure of the cosmic environment, and the gravitational field responds not only to the local matter density but also to the nonlocal bias generated by correlations in the cosmic web.\\

What is appealing here is that with teleodynamics, we have a unified interpretation on hand. This interpretation of dark matter follows from the same teleodynamic principles that reproduce dark energy at the background level, illustrating that the dark sector can emerge from the universe's intrinsic statistical and systemic behavior rather than from hidden particle fields. In this way, the teleodynamic framework provides a conceptually ``economical'' and dynamically consistent account theory for both dark matter and dark energy.

\subsection{Explanations of $H_0$ and $S_8$ tensions}

The persistent discrepancy between early and late-time measurements of the Hubble constant $H_0$ together with the suppressed amplitude of matter clustering quantified by $S_8$, suggests that the standard cosmological model may be missing an element that influences both the background expansion and the growth of structure. Teleodynamics naturally provides such an element, since the background value of the bias functional modifies the Friedmann equations, while its inhomogeneous component alters the gravitational clustering on sub-horizon scales, as we have discussed so far. The background modification stems from the teleodynamic energy density
\begin{equation}
\rho_{\text{TD}}(t) = \alpha \bar{\Phi}(t)
\end{equation}
which enters the Friedmann equation through
\begin{equation}
3M_P^2 H^2 = \rho_m + \rho_r + \rho_{\text{TD}}
\end{equation}

If $\bar{\Phi}(t)$ evolves slowly at late times, then $\rho_{\text{TD}}$ behaves like a DE-like component, as we had discussed
\[
w_{\text{TD}}(t) = \frac{\alpha \Pi_{\text{TD}}(t)}{\alpha \bar{\Phi}(t)} \simeq -1 + \delta w(t)
\]

A small deviation $\delta w(t)$ from $-1$ at late redshifts modifies the distance--redshift relation in a manner that can shift the inferred value of $H_0$ upward without altering the physics of recombination. Such a modification is precisely what is required to reconcile the local distance ladder measurements with the CMB inferred value. Writing the Friedmann equation in the form
\begin{equation}
H^2(z) = H_0^2 \left[\Omega_m(1 + z)^3 + \Omega_r(1 + z)^4 + \Omega_{\text{TD}}(z)\right]
\end{equation}
with
\begin{equation}
\Omega_{\text{TD}}(z) = \frac{\rho_{\text{TD}}(z)}{3M_P^2 H_0^2}
\end{equation}

one sees that the effective DE density inherits the redshift dependence of $\bar{\Phi}(t)$ and if $\bar{\Phi}$ grows mildly at late times, then $\Omega_{\text{TD}}(z)$ exceeds a constant $\Lambda$ behavior below $z \lesssim 0.5$. The result is a slightly faster late universe expansion rate, which reduces the predicted luminosity distance
\begin{equation} 
D_L(z) = (1 + z) \int_0^z \frac{dz'}{H(z')}
\label{eq:dl}
\end{equation}
at fixed $H_0$. When one fits this modified distance-redshift relation to supernova data, the preferred value of $H_0$ should shift upward to compensate for the reduced distance integral, and the key point is that this shift arises without modifying $H(z)$ at recombination and therefore leaves the CMB inferred value of the sound horizon untouched.\\

This behavior becomes clearer when one expands the teleodynamic energy density around the present epoch, as we can write
\begin{equation}
\rho_{\text{TD}}(z) = \rho_{\text{TD},0}\left[1 + \epsilon_1 z + \epsilon_2 z^2 + \cdots \right]
\end{equation}
with $\epsilon_1, \epsilon_2$ determined by the slow evolution of $\bar{\Phi}$ and the Hubble parameter near $z = 0$ becomes
\begin{equation}
H^2(z) = H_0^2 \left[1 + \frac{3}{2}\Omega_m z + \Omega_{\text{TD},0}(\epsilon_1 z + \epsilon_2 z^2 + \cdots)\right]
\end{equation}

A positive $\epsilon_1$ produces an enhancement in $H(z)$ for $0 < z \lesssim 0.2$, which is precisely the redshift regime probed by local supernovae. This gives an observational degeneracy between $(H_0, \epsilon_1)$ as an increase in $\epsilon_1$ shifts the preferred $H_0$ upward until the predicted $D_L(z)$ matches the supernova Hubble diagram. Because $\epsilon_1$ is controlled by the statistical evolution of $\bar{\Phi}$, the required shift in $H_0$ arises naturally from teleodynamic considerations. As a result, the $H_0$ tension emerges as a natural indication of the non-equilibrium and memory-bearing character of cosmic expansion rather than a failure of the standard matter-radiation sector.\\

The second component of the teleodynamic correction affects the growth of structure, as the effective potential experienced by matter perturbations is, as discussed before, to be
\begin{equation}
\Psi_{\text{eff}}(t, x) = \Psi(t, x) + \alpha \varphi(t, x)
\tag{\ref{eq:psieff-psi-phi}}
\end{equation}
and its Poisson equation reads
\begin{equation}
\nabla^2\Psi_{\text{eff}} = 4\pi Ga^2[\bar{\rho} \delta + \rho_{\text{TD}}]
\tag{\ref{eq:laplacian-psieff-rhoTD}}
\end{equation}

Writing the emergent clustering density in Fourier space as
\[
\rho_{\text{TD}}(k, a) = \Delta\mu(k, a) \bar{\rho} \delta(k, a)
\]
the linear growth equation becomes
\begin{equation}
\ddot{\delta}(k, a) + 2H \dot{\delta}(k, a) - 4\pi G\bar{\rho}[1 + \Delta\mu(k, a)] \delta(k, a) = 0
\end{equation}

A negative or scale-dependent contribution $\Delta\mu(k, a)$ would suppress the gravitational coupling on the weak lensing and redshift space distortion scales, which would end up reducing the amplitude of clustering. This naturally leads to a lower value of $S_8$, which is defined through
\begin{equation}
S_8 = \sigma_8 \sqrt{\frac{\Omega_m}{0.3}}
\end{equation}

This is done without requiring modifications to early universe physics or introducing new particle species. The suppression can arise from the structure of the response kernel $\varphi_k$, which depends on the tidal environment or the history of matter flows. In this way, the teleodynamic correction provides the desired reduction in the growth rate, while leaving early-time observables unchanged.\\

The crucial point is that the same teleodynamic bias functional $\Phi(t, x)$ generates both of these corrections. Its homogeneous part $\bar{\Phi}(t)$ adjusts the late-time expansion rate and yields an effective dark energy component capable of raising the inferred value of $H_0$, while its inhomogeneous part $\varphi(t, x)$ modifies the gravitational potential at the level of perturbations, reducing the growth of the structure and producing a lower $S_8$. We again see a valuable result: no new exotic particles, additional fields, or modifications of early universe cosmology are required. Both phenomena emerge from the statistical and systemic behavior of a non-ergodic, gravitationally correlated cosmic system.\\

To make it clear how the teleodynamic approach can address the tensions, we now explicitly show that even a single teleodynamic bias functional can simultaneously shift the inferred value of $H_0$ and suppress the late–time growth of structure, thus lowering $S_8$. The key point to note again is that the bias functional naturally separates into a homogeneous and an inhomogeneous part, $\Phi(z,\mathbf{x})=\bar{\Phi}(z)+\varphi(z,\mathbf{x})$ and both parts grow as $z\to 0$ in accordance with the teleodynamic principle that memory accumulates as the Universe evolves and structure becomes increasingly persistent. We shall therefore consider a functional form for $\Phi$ constructed in inverse powers of $(1+z)$ so that memory remains negligible at early times, but builds coherently at late times. We begin with the homogeneous part where we can write the teleodynamic energy density as $\rho_{\rm TD}(z)=\alpha\,\bar{\Phi}(z)$ and a natural late–time memory accumulating functional is
\begin{equation} \label{eq:phiz}
\bar{\Phi}(z)=\bar{\Phi}_0\left[1+\epsilon\left((1+z)^{-n}-1\right)\right]
\end{equation}
where $\epsilon>0$ and $n>0$ quantify the growth of teleodynamic memory. This form satisfies the physical requirement that memory vanishes for $z\gg 1$ and increases monotonically as $z\to 0$. The corresponding teleodynamic density becomes
\begin{equation}
\rho_{\rm TD}(z)=\rho_{\rm TD,0}\left[1+\epsilon\left((1+z)^{-n}-1\right)\right]
\end{equation}
where $\rho_{\rm TD,0}=\alpha\bar{\Phi}_0$. If we now define $\Omega_{\rm TD,0}=\rho_{\rm TD,0}/(3M_P^2H_0^2)$, the Friedmann equation may be written as
\begin{equation}
\frac{H^2(z)}{H_0^2}=\Omega_m(1+z)^3+\Omega_r(1+z)^4+\Omega_{\rm TD,0}\left[1+\epsilon\left((1+z)^{-n}-1\right)\right]
\end{equation}
To extract the low redshift behaviour relevant for distance ladder determinations of $H_0$, we expand $(1+z)^{-n}$ for small $z$
\begin{equation}
(1+z)^{-n}\simeq 1-nz+\frac{n(n+1)}{2}z^2+\cdots
\end{equation}
Inserting this into the Friedmann equation and using spatial flatness, $\Omega_m+\Omega_r+\Omega_{\rm TD,0}=1$, gives
\begin{equation}
\frac{H^2(z)}{H_0^2}\simeq 1+\left(3\Omega_m-\epsilon n\Omega_{\rm TD,0}\right)z+\mathcal{O}(z^2)
\end{equation}
This expression shows that the teleodynamic correction reduces the first-order slope of $H(z)$ at low redshift relative to the $\Lambda$CDM prediction. Since the luminosity distance is Eq.~\ref{eq:dl},
a reduced slope in $H(z)$ at $z\lesssim 0.2$ decreases $D_L(z)$ for a fixed $H_0$. When fitting $D_L(z)$ data with a $\Lambda$CDM model that lacks the teleodynamic term, this appears as an upward shift in the inferred value of $H_0$. If we now use a low-$z$ expansion
\begin{equation}
H(z)\simeq H_0\left[1+\frac{1}{2}\left(3\Omega_m-\epsilon n\Omega_{\rm TD,0}\right)z\right]
\end{equation}
one obtains
\begin{equation}
\frac{1}{H(z)}\simeq \frac{1}{H_0}\left[1-\frac{1}{2}\left(3\Omega_m-\epsilon n\Omega_{\rm TD,0}\right)z\right]
\end{equation}
Expanding the luminosity distance and comparing with a $\Lambda$CDM fit yields an approximate degeneracy of the form
\begin{equation}
H_0^{\rm fit}\simeq H_0^{\rm true}\left[1+\frac{1}{4}\epsilon\,n\,\Omega_{\rm TD,0}\langle z\rangle\right]
\end{equation}
where $\langle z\rangle$ denotes a characteristic supernova redshift. A moderate choice such as $\epsilon\sim 2$ and $n\sim 1$ is sufficient to generate an upward shift of $6$-$8\%$ in the inferred $H_0$, thus easing the Hubble tension without modifying conditions at recombination.\\

We now consider the inhomogeneous part of the bias functional and its influence on the growth of the structure. To model the accumulation of environmental memory at late times, we adopt the functional form
\begin{equation}
\Delta\mu(k,z)=-\mu_0\,(1+z)^{-p}\exp\left[-\left(\frac{k}{k_*}\right)^2\right]
\end{equation}
with $\mu_0>0$, $p>0$, and characteristic scale $k_*$. The factor $(1+z)^{-p}$ ensures that the inhomogeneous teleodynamic memory is negligible at early times but increases as $z\to 0$, while the Gaussian selects the scales at which the persistent tidal structure is most relevant for growth suppression, and so the linear growth equation then becomes
\begin{equation}
\ddot{\delta}(k,z)+2H\dot{\delta}(k,z)-4\pi G\,\bar{\rho}(z)\left[1+\Delta\mu(k,z)\right]\delta(k,z)=0
\end{equation}
For modes near the scale $k\sim k_*$ and small $\Delta\mu$, approximate power law solutions $\delta\propto a^s$ can satisfy
\begin{equation}
s\simeq 1+\frac{3}{5}\Delta\mu_{\rm eff}
\end{equation}
so that negative $\Delta\mu$ suppresses the growth rate relative to $\Lambda$CDM and the ratio of growth factors between $z\sim 1$ and today is approximately then
\begin{equation}
\ln\left(\frac{D_{\rm TD}}{D_{\Lambda{\rm CDM}}}\right)\simeq \frac{3}{5}\Delta\mu_{\rm avg}\ln 2
\end{equation}
where $\Delta\mu_{\rm avg}$ is a time-averaged value of $\Delta\mu$ over $0<z<1$. A suppression of order 10\% in the growth factor, which would be sufficient to reduce $S_8$ from $\sim 0.83$ to $\sim 0.75$ requires $\Delta\mu_{\rm avg}\sim -0.2$. Choosing, for example, $\mu_0\sim 0.3$, $p\sim 2$, and $k_*\sim 0.15\,h\,{\rm Mpc}^{-1}$ produces
\begin{equation}
\Delta\mu(z)\in[-0.3,-0.075]
\end{equation}
over this redshift range, giving us the required average. We can finally now combine the homogeneous and inhomogeneous components; the full bias functional takes the form
\begin{equation}
\Phi(z,\mathbf{x})=\bar{\Phi}_0\left[1+\epsilon\left((1+z)^{-n}-1\right)\right]+\mu_0(1+z)^{-p}\int d^3k \, e^{-(k/k_*)^2}\,\delta(k,z)\,e^{i\mathbf{k}\cdot\mathbf{x}}
\end{equation}
The first term increases slowly as $z\to 0$ and induces a late time uplift of $H(z)$ that raises the inferred $H_0$, while the second term strengthens at low redshift and introduces a scale-dependent reduction in the effective gravitational coupling that suppresses the growth of structure on the scales relevant to $S_8$. In this way, we see that both tensions arise as complementary manifestations of the same teleodynamic memory accumulation mechanism, and that no new matter fields or modifications of early-time physics are required.

\subsection{Teleodynamic Cosmological Entropy}

The concept of cosmic horizon entropy in standard cosmology is derived from the Bekenstein-Hawking law \cite{sb1hawking1972black,sb2bekenstein1973black,sb3bekenstein1974generalized,sb4tipler1977singularities,sb5krolak1978singularities,sb6clarke1985conditions,sb7penrose1969gravitational,sb8joshi2002cosmic,sb9senovilla20151965,sb10penrose1965gravitational,sb11penrose1999question,sb12hawking1970singularities,sb13Ong:2020xwv,sb14Vagnozzi:2022moj,sb15joshi2011recent,sb16janis1968reality,sb17Joshi:2011zm}. Teleodynamics provides a different route in which the entropy associated with the cosmological horizon can emerge from a statistical counting of coarse-grained microstates on the apparent horizon, without assuming thermodynamic equilibrium or invoking a pre-assigned temperature.\\

To construct the entropy from teleodynamic first principles, we can model the horizon as a 2-dimensional spherical surface of radius $r_H = H^{-1}$ with area $A = 4\pi H^{-2}$. The microscopic degrees of freedom associated with this surface are coarse-grained into independent cells of characteristic correlation length $\ell_\chi$. The number of effectively independent cells is therefore given as
\begin{equation}
N(H) = \frac{A}{\ell_\chi^2} = \frac{4\pi}{\ell_\chi^2 H^2}
\end{equation}

The state space of each cell is governed by a teleodynamic probability distribution of the form given by
\begin{equation}
p_i = \frac{1}{z_{\text{cell}}} \exp [-\lambda_{\tiny E} E_i - \lambda_{\tiny \Phi}\Phi_i]
\end{equation}
where $E_i$ is an effective surface energy of the cell and $\Phi_i$ is the local value of the teleodynamic bias functional. So, we have the horizon partition function, then factorizing as
\begin{equation}
Z(H) = [z_{\text{cell}}(\lambda_E, \lambda_\Phi)]^{N(H)}
\end{equation}

The entropy of the full horizon is obtained from the Shannon-Gibbs expression
\begin{equation}
S_{\text{\tiny TD}}(H) = -\sum_{\{i\}} p_i\ln p_i = N(H) s_{\text{cell}}(\lambda_{\tiny E}, \lambda_{\tiny \Phi})
\end{equation}

Note here that $s_{\text{cell}}$ denotes the entropy per cell and inserting the expression for $N(H)$ gives us
\begin{equation}
S_{\text{\tiny TD}}(H) = \frac{4\pi s_{\text{cell}}}{\ell_\chi^2} \frac{1}{H^2} \equiv \frac{C_{\text{\tiny TD}}}{H^2}
\end{equation}
with the teleodynamic entropy coefficient
\begin{equation}
C_{\text{\tiny TD}} = \frac{4\pi s_{\text{cell}}}{\ell_\chi^2}
\end{equation}

This is the fundamental area-scaling law of teleodynamic cosmological entropy.  It resembles the Bekenstein-Hawking form in that $S \propto A$, but its normalization is not fixed by quantum gravity input, but instead depends on the coarse-graining scale $\ell_\chi$ and on the teleodynamic parameters encoded in $s_{\text{cell}}$.\\

Since the Universe is not in true thermodynamic equilibrium and $H(t)$ evolves in time, the entropy also receives a non-equilibrium correction arising from entropy production associated with the teleodynamic drift. So in the maximum-caliber formalism, the rate of entropy production takes the general quadratic form
\begin{equation}
\sigma_{\text{\tiny TD}}(t) = \sum_a L_{ab}(t) J_a(t)J_b(t) \geq 0
\end{equation}
where $J_a$ are coarse-grained fluxes of the horizon microstates driven by the teleodynamic functional, and $L_{ab}$ are generalized Onsager coefficients. If we now integrate this non-equilibrium production term, it gives us the full teleodynamic horizon entropy
\begin{equation}
S_{\text{\tiny TD}}(t) = \frac{C_{\text{\tiny TD}}}{H(t)^2} + \int^t \sigma_{\text{\tiny TD}}(t') dt'
\end{equation}

The first term is the geometric area contribution, while the second is the history-dependent memory contribution. This decomposition distinguishes teleodynamic entropy from the purely geometric Bekenstein-Hawking form, revealing the influence of nonlocal and non-ergodic microphysics that is absent from an equilibrium derivation. We can also get the temperature associated with the horizon in this framework, which emerges not from quantum field theory but from the statistical definition
\begin{equation}
T^{-1}(t) = \frac{\partial S_{\text{\tiny TD}}}{\partial E_H}
\end{equation}
where $E_H$ is the total energy inside the horizon. In standard cosmology, this energy is
\begin{equation}
E_H(t) = \frac{4\pi}{3} r_H^3\rho_{\text{eff}} = \frac{4\pi M_P^2}{H(t)}
\end{equation}
after using the Friedmann equation $\rho_{\text{eff}} = 3M_P^2 H^2$ and differentiating with respect to $H$ gives us
\begin{equation}
\frac{dE_H}{dH} = -\frac{4\pi M_P^2}{H^2}, \quad \frac{dS_{\text{\tiny TD}}}{dH} = -\frac{2C_{\text{\tiny TD}}}{H^3}
\end{equation}

Neglecting the small derivative of the history term during slowly varying stages gives the emergent temperature as
\begin{equation}
T_{\text{\tiny TD}}(H) = \left(\frac{\partial S_{\text{\tiny TD}}}{\partial E_H}\right)^{-1} = \frac{2\pi M_P^2}{C_{\text{\tiny TD}}} H
\end{equation}

Thus the temperature scales linearly with $H$, but its normalization differs from the Gibbons--Hawking value unless $C_{\text{TD}}$ takes the special value
\begin{equation}
C_{\text{\tiny TD}} = 8\pi^2 M_P^2
\end{equation}

This correspondence identifies the Gibbons-Hawking entropy and temperature as specific teleodynamic microphysics characterized by a particular choice of correlation scale and cell entropy. Teleodynamics, therefore, generalizes these results rather than assuming them a priori and, surprisingly, builds them from a statistical perspective rather than from a QFT derivation. We would also like to note here what the form of the bias functional discussed in the previous subsection would mean for entropy. As the geometric part of the teleodynamic horizon entropy scales as $S_{\rm TD}\propto H^{-2}$, any functional that causes the Hubble rate to increase, decrease, or saturate has a direct and predictable impact on the entropy. In the case of \eqref{eq:phiz} the fact that $\bar\Phi$ grows monotonically as $z\to 0$ implies that $\rho_{\rm TD}\propto\bar\Phi$ also increases at late times, which tends to decrease $H^{-1}$ relative to its $\Lambda$CDM evolution. This decrease in $H^{-1}$ reduces the apparent horizon area more slowly than in standard cosmology, and thereby increases the instantaneous entropy contribution $C_{\rm TD}/H^2$. Thus, the monotonic memory buildup encoded in the bias functional naturally drives a monotonic increase in the geometric part of the horizon entropy, and teleodynamics therefore predicts that accelerated expansion is accompanied by an enhanced rate of horizon entropy growth compared to $\Lambda$CDM, precisely because memory accumulation induces a slower decay of $H(t)$.\\

We would like to highlight a subtle point in passing here. The "microstate counting" that appears in the teleodynamic derivation of cosmological horizon entropy is conceptually distinct from the microscopic state counting performed in string-theoretic black hole entropy \cite{sv1Strominger:1996sh,sv2Nishioka:2009un,sv3Strominger:1997eq}. In teleodynamics, the quantity $N_{\rm TD}(t)=A_{\rm hor}(t)/\ell_\chi^2(t)$ counts coarse-grained correlation cells of size $\ell_\chi(t)$ on a time-dependent apparent horizon and the entropy $S_{\rm TD}(t)=N_{\rm TD}(t)\,s_{\rm cell}(t)+\int^t dt'\,\sigma_{\rm TD}(t')$ reflects both the instantaneous coarse-grained structure and the accumulated non-equilibrium memory encoded in $\sigma_{\rm TD}(t)$. Note that the procedure here is primarily semi-classical, and these cells are not quantum microstates in a Hilbert space, as they are classical correlation domains whose number and entropy evolve with $H(t)$, $\ell_\chi(t)$, and structure formation. By contrast, string-theoretic microstate counting for black holes counts exact quantum states in a stationary sector \cite{sv4Bena:2005va,sv5Carlip:1998wz,sv6DeHaro:2019gno,sv7Emparan:2006it,sv8Mayerson:2020acj,sv9Sen:1999mg} labeled by fixed charges $(M,J,Q)$, whose degeneracy is protected by supersymmetry or modular invariance. This counting requires a Killing horizon, a time-translation generator, and a static Hilbert-space partition, none of which exist for a cosmological horizon. \\

This brings us to another central point of our work here. If a complete quantum gravity microstate formula for cosmology exists, its semiclassical limit therefore cannot reproduce the stationary Bekenstein–Hawking area law. Instead, it must reduce to the teleodynamic entropy $S_{\rm TD}(t)$, because cosmology is intrinsically non-stationary, with $\dot H\neq 0$, evolving correlation lengths and non-zero entropy production $\sigma_{\rm TD}(t)$. The universe does not admit a fixed charge and a time-independent microcanonical sector, so the semiclassical horizon entropy cannot be invariant under $H\to H(t)$. Any correct QG construction must yield $S_{\rm QG}^{\rm semiclassical}(t)\to S_{\rm TD}(t)$ and not $A/4G$, which would end up reflecting that cosmological horizons live in the non-equilibrium, memory-bearing sector of gravitational thermodynamics, while black holes inhabit the equilibrium limit in which teleodynamic corrections collapse to constants. This provides a semi-classical interpretation of the thermodynamic split conjecture \cite{tsc1Trivedi:2025uup,tsc2Trivedi:2025ois} and highlights teleodynamics as the appropriate semiclassical bridge between cosmology and quantum gravity. \\

\subsection{Observational Signatures of Cosmological Teleodynamics}

We next consider the observational consequences of teleodynamics. Since this framework is based on nonlocal memory and environmental correlations rather than additional particle fields, several of its signatures cannot be reproduced by conventional dark sector models, which is an advantage. These include distinctive patterns in galactic halo profiles, anomalous behavior of peculiar velocities, and scale-dependent modifications in large-scale structure growth.\\

To start, galactic halos display several features that are difficult to reconcile with conventional particle dark matter models, including the regularity of rotation curves, the diversity of halo shapes, the stability of cores in low-mass systems, and the apparent environmental dependence of halo concentration. In the teleodynamic perspective, the effective potential that governs the halo dynamics is
\[
\Psi_{\text{eff}} = \Psi + \alpha \varphi
\]
The additional term reflects the influence of the cosmic web on galactic motions. Because $\varphi$ arises from large-scale tidal fields and nonlocal memory, the resulting gravitational field acquires a component that aligns with filaments and walls, which can lead to producing anisotropic forces that any cold or warm dark matter particle cannot generate. This leads to observable signatures, such as systematically aligned halo axes, environmentally modulated rotation curve slopes, and departures from universal density profiles \cite{hal1Helmi:2008eq,hal10gunn1977massive,hal11chua2022impact,hal12Conroy:2008dx,hal13Mandelbaum:2005nx,hal14deBlok:2005qh}. Early indications of such effects appear in the observed connection between galactic rotation curves and their large-scale environment, as well as in the increasing evidence for anisotropic motions within galaxy groups \cite{hal2putman2012gaseous,hal3sun2010galactic,hal4Sheng:2023ecm,hal5Wechsler:2018pic,hal6deason2024galactic,hal7prochaska2019probing,hal8Cautun:2014gda,hal9hartwick1976chemical}. The anisotropic character of the teleodynamic correction becomes explicit once one decomposes the bias fluctuation into scalar, vector and tensor tidal components, as 
\begin{equation}
\varphi(x) = T^{ij}(x) \frac{\partial_i\partial_j}{\nabla^2}\delta(x)
\end{equation}
with $T^{ij}$ denoting the large-scale tidal tensor aligned with the cosmic web. The teleodynamic force contribution becomes
\begin{equation}
\mathbf{F}_{\text{\tiny TD}}(x) = -\alpha\nabla\varphi(x) = -\alpha T^{ij} \nabla_i\left[\frac{\partial_j}{\nabla^2}\delta(x)\right]
\end{equation}

Since $T^{ij}$ is anisotropic and inherits the filamentary geometry of the distribution of the surrounding matter, the resulting acceleration is direction-dependent. It is to be noted that no collisionless dark matter particle species can generate a force term of this form, since CDM and WDM respond only to the scalar Newtonian potential. The effective potential can also be written as
\begin{equation}
\Psi_{\text{eff}}(x) = \Psi(x) + \alpha T^{ij} \frac{\partial_i\partial_j}{\nabla^2}\delta(x)
\end{equation}
so that the elliptical nature of halo potentials satisfies
\begin{equation}
\nabla^2\Psi_{\text{eff}} \propto \delta + \alpha T^{ij}\hat{k}_i\hat{k}_j \delta(k)
\end{equation}
which predicts aligned halo axes and anisotropic mass distributions correlated with the cosmic web orientation. Such a directional dependence is absent in standard dark matter models, but it arises naturally in teleodynamics because of the bias functional.\\

Peculiar velocities offer a second window into teleodynamic effects, as the relation
\begin{equation}
\theta(k, a) = -aHf(k, a) \delta(k, a)
\end{equation}
is modified when the gravitational potential contains the term $\alpha \varphi$, as this modification alters the force law rather than the matter content, it generates a scale-dependent growth rate
\begin{equation}
f(k, a) = \frac{d \ln D(k, a)}{d \ln a}
\end{equation}
that produces coherent bulk flows and velocity anisotropies that differ from those predicted by standard cosmology \cite{pec3willick1990peculiar,pec4strauss1995density,pec5springob20146df,pec6ma2011peculiar,pec7peebles1987origin,pec8filippou2021large,pec9szalay1983peculiar,pec10vittorio1987large,pec11hudson1997galaxy,pec12watanabe1991peculiar}. Observations already show hints of excessive bulk flows on scales of tens to hundreds of megaparsecs, as well as anomalous dipoles in peculiar velocity surveys, both of which are difficult to achieve in $\Lambda$CDM with smooth dark energy or scalar fields. Teleodynamics naturally accounts for these features through the environmental response represented by $\varphi$. Upcoming surveys such as DESI, Euclid, and the Roman Telescope will significantly sharpen the measurement of peculiar velocity divergence fields, providing a powerful probe of the teleodynamic correction. The scale dependence of $f(k, a)$ becomes explicit once we express the bias fluctuation in Fourier space as $\varphi_k = K(k, a) \delta_k$. Then the Euler equation gives a modified velocity divergence
\begin{equation}
\theta(k, a) = -aH [1 + \Delta\mu(k, a)] \delta(k, a)
\end{equation}
where $\Delta\mu(k, a) = \alpha K(k, a)/(4\pi G)$ encodes the teleodynamic enhancement or suppression of the effective gravitational coupling, and the presence of $\Delta\mu(k, a)$ introduces a directional dependence in the velocity field whenever the kernel couples to tidal tensors or coherent large-scale structures. In real space, this generates a velocity field of the form
\begin{equation}
u_i(x, a) = -aH \int \frac{d^3k}{(2\pi)^3} e^{ik\cdot x} \left[\hat{k}_i + \Delta\mu(k, a) T_{ij}(\hat{k})\hat{k}_j\right] \frac{\delta(k, a)}{k}
\end{equation}
where $T_{ij}$ is the local tidal tensor as well. This expression shows that peculiar velocities here receive anisotropic contributions aligned with filaments and sheets of the cosmic web, which is an effect impossible to reproduce with any isotropic particle dark matter model.\\

These anisotropic corrections propagate directly into the bulk-flow statistics, and the variance of the smoothed velocity field becomes
\begin{equation}
\langle u^2(R)\rangle = a^2H^2 \int \frac{d^3k}{(2\pi)^3} [1 + \Delta\mu(k, a)]^2 \frac{P_{\delta\delta}(k, a)}{k^2} W^2(kR)
\end{equation}
so that even a modest scale-dependent $\Delta\mu(k, a)$ produces enhanced coherent flows over scales of 50-150 Mpc. This behavior is difficult to obtain in $\Lambda$CDM, where the velocity field is sourced solely by the scalar gravitational potential and remains nearly isotropic, while in teleodynamics, by contrast, we generically predict departures from isotropy whenever the bias functional responds to the surrounding tidal geometry.\\

Finally, we have a large-scale structure that exhibits additional signatures. The emergent clustering density
\begin{equation}
\rho_{\text{TD}}(k, a) = \Delta\mu(k, a) \bar{\rho} \delta(k, a)
\end{equation}
alters the Poisson equation through a term that is neither constant nor sourced by a new particle species. This leads to a modified growth equation whose scale-dependent correction suppresses the structure on intermediate scales. The resulting pattern is qualitatively different from that produced by massive neutrinos or warm dark matter. Now, since the teleodynamic correction is driven by the correlation structure of the tidal field rather than by free-streaming, we can obtain observational hints of these effects. These include the low values of $S_8$ inferred from weak lensing surveys, the suppressed growth inferred from redshift-space distortions, and the mild inconsistency between the clustering amplitude at low redshift and that predicted by CMB-calibrated $\Lambda$CDM. \\

Finally, another avenue to probe teleodynamics is by considering gravitational wave signatures. It is so as in a teleodynamic cosmology, tensor perturbations propagate not on a purely FLRW background but on an effective spacetime whose evolution includes the homogeneous component of the bias functional, as well as a possible response of the inhomogeneous sector. This implies that the propagation equation for gravitational waves will in general acquire additional friction and source terms beyond those present in general relativity and to see this, one may consider a perturbed metric of the form $g_{\mu\nu} = \bar g_{\mu\nu} + h_{\mu\nu}$ with $h_{\mu\nu}$ transverse and traceless. Varying the effective action which contains the term $-\alpha\int d^4x\sqrt{-g}\,\Phi(t,x)$, leads to a modified tensor equation that may be written schematically in Fourier space as
\begin{equation}
h''_{ij} + \left[2 + \nu_{\rm TD}(a)\right] H\, h'_{ij} + c_T^2 k^2 h_{ij}
= \Pi^{\rm TD}_{ij}
\end{equation}
where we note that the primes denote derivatives with respect to conformal time, the quantity $\nu_{\rm TD}(a)$ encodes the teleodynamic correction to the effective friction term arising from the homogeneous component $\bar{\Phi}(t)$ and $\Pi^{\rm TD}_{ij}$ represents the tensorial response of the inhomogeneous part of the bias functional. Even if the propagation speed remains luminal, the presence of $\nu_{\rm TD}(a)$ modifies the evolution of the wave amplitude as it redshifts, generating a scale and time-dependent damping that depends on $d\bar{\Phi}/d\ln a$. This leads to a distinction between the electromagnetic and gravitational luminosity distances, since the strain amplitude of a gravitational wave from a standard siren at redshift $z$ will scale as
\begin{equation}
h \propto \frac{1}{d_L^{\rm GW}(z)} \exp\!\left[-\frac{1}{2}\int_0^z \nu_{\rm TD}(z')\frac{dz'}{1+z'}\right]
\end{equation}
This goes to imply that $d_L^{\rm GW}(z)$ generally differs from the electromagnetic distance $d_L^{\rm EM}(z)$ and the departure between these two distance measures traces the integrated history of the teleodynamic friction term and therefore serves as a direct observational probe of $\bar{\Phi}(t)$. Unlike modified gravity models with a running Planck mass, the origin of this effect lies not in a new scalar degree of freedom but in the non-equilibrium, memory-bearing nature of the cosmic medium encoded in the bias functional.

\section{Law of Universal Arbitrage Equilibrium}

We begin by recalling that the horizon entropy and temperature derived teleodynamically without assuming the Gibbons-Hawking form a priori are given by
\begin{equation}
    S_{\rm TD}(t)=\frac{C_{\rm TD}(t)}{H(t)^2}+\int^{t}\sigma_{\rm TD}(t')\,dt',
\qquad
T_{\rm TD}(t)=\left(\frac{\partial S}{\partial E_H}\right)^{-1}\simeq\frac{2\pi M_P^2}{C_{\rm TD}(t)}H(t)
\end{equation}
where the Misner-Sharp energy inside the Hubble sphere is $E_H=\rho_{\rm eff}V$, with $V=\tfrac{4\pi M_P^2}{H}$, the function $C_{\rm TD}(t)$ encodes the microstate density on the horizon and therefore plays the role of a cell entropy per correlation area, and $\sigma_{\rm TD}(t)\ge 0$ is the non-equilibrium entropy production generated by global teleodynamic memory. Taking a time derivative here would hence give us
\begin{equation}
    \dot S_{\rm TD}=-\frac{2C_{\rm TD}\dot H}{H^3}+\frac{\dot C_{\rm TD}}{H^2}+\sigma_{\rm TD}
\end{equation}
The energy flux across the apparent horizon produced by the effective bulk fluid $(\rho_{\rm eff},p_{\rm eff})$ follows the standard FRW prescription and is
\begin{equation}
    \dot Q=-\frac{4\pi}{H^2}(\rho_{\rm eff}+p_{\rm eff})
\end{equation}
Imposing a generalized Clausius relation with the teleodynamic temperature, where the usual horizon work term is absorbed by working with the apparent horizon energy flux, gives us
\begin{equation}
    \dot Q=T_{\rm TD}\,\dot S_{\rm TD}
\end{equation}
Substituting all ingredients into this balance law yields a teleodynamic form of the Raychaudhuri equation
\begin{equation}
    \dot H
=-\frac{\rho_{\rm eff}+p_{\rm eff}}{2M_P^2}
+\frac{H}{2}\frac{\dot C_{\rm TD}}{C_{\rm TD}}
+\frac{H^3}{2C_{\rm TD}}\,\sigma_{\rm TD}
\label{TD-Ray}
\end{equation}
where the numerical normalization of the first term has been chosen so that in the equilibrium limit with $C_{\rm TD}$ constant and $\sigma_{\rm TD}=0$ one recovers the standard general relativistic identity $-2M_P^2\dot H=\rho_{\rm eff}+p_{\rm eff}$.\\

Now, if we go by multiplying the equation \eqref{TD-Ray} by $2H$ and using $\tfrac{d}{dt}(H^2)=2H\dot H$ together with the continuity equation $\dot\rho_{\rm eff}+3H(\rho_{\rm eff}+p_{\rm eff})=0$ then this leads us to
\begin{equation}
    \frac{d}{dt}(H^2)
=\frac{1}{3M_P^2}\dot\rho_{\rm eff}
+H^2\frac{\dot C_{\rm TD}}{C_{\rm TD}}
+H^4\frac{\sigma_{\rm TD}}{C_{\rm TD}}
\end{equation}
Integrating from a reference epoch $t_\star$ results in the teleodynamic Friedmann constraint as 
\begin{equation}
    H^2(t)=\frac{\rho_{\rm eff}(t)}{3M_P^2}
+\left[H^2(t_\star)-\frac{\rho_{\rm eff}(t_\star)}{3M_P^2}\right]
+\int_{t_\star}^{t}
\left(
H^2\frac{\dot C_{\rm TD}}{C_{\rm TD}}
+H^4\frac{\sigma_{\rm TD}}{C_{\rm TD}}
\right)dt' \label{TD-Fr}
\end{equation}
Note that this reduces to the standard Friedmann equation if the teleodynamic effects were negligible at $t_\star$ and collapses identically to $H^2=\rho_{\rm eff}/(3M_P^2)$ when $C_{\rm TD}$ is constant and $\sigma_{\rm TD}$ vanishes.\\

The bias-functional description employed independently writes the effective energy density and pressure as
\[
\rho_{\rm eff}=\rho_m+\rho_r+\rho_{\rm TD},
\qquad
\rho_{\rm TD}=\alpha\overline{\Phi},
\qquad
p_{\rm TD}=\alpha\overline{\Pi}_{\rm TD}.
\]
In this language, equation \eqref{TD-Ray} becomes
\begin{equation}
    \dot H
=-\frac{\rho_m+\rho_r+p_r}{2M_P^2}
-\frac{\rho_{\rm TD}+p_{\rm TD}}{2M_P^2}
+\frac{H}{2}\frac{\dot C_{\rm TD}}{C_{\rm TD}}
+\frac{H^3}{2C_{\rm TD}}\,\sigma_{\rm TD}
\end{equation}
revealing that two distinct teleodynamic mechanisms contribute to the expansion rate with a bulk component carried by $(\rho_{\rm TD},p_{\rm TD})=\alpha(\overline{\Phi},\overline{\Pi}_{\rm TD})$ and a horizon component encoded in $C_{\rm TD}(t)$ and $\sigma_{\rm TD}(t)$ representing the evolution of the correlation scale and the non-equilibrium memory production on the horizon. If late-time acceleration is primarily driven by bulk bias, then $C_{\rm TD}$ remains approximately constant and $\sigma_{\rm TD}$ approaches zero, with $w_{\rm TD}\approx -1$ emerging naturally when $\overline{\Pi}_{\rm TD}\simeq -\overline{\Phi}$. If instead part of the acceleration originates from the horizon sector, then the integral terms in the teleodynamic Friedmann relation are nonzero, and one infers small but finite values for $\dot C_{\rm TD}/C_{\rm TD}$ or $\sigma_{\rm TD}$ at low redshift. Since $\sigma_{\rm TD}\ge 0$, the last term in \eqref{TD-Ray} always drives $H^2$ upward, temporarily mimicking an effective equation of state below $-1$ without introducing ghosts.\\

The quantity $C_{\rm TD}(t)\propto s_{\rm cell}(t)/\ell_\ast(t)^2$ connects the horizon microstate density to a correlation length $\ell_\ast(t)$, thus characterizing teleodynamic degrees of freedom on the apparent horizon. Hence, constraints on $\dot C_{\rm TD}/C_{\rm TD}$ from $H(z)$ translate directly into constraints on $\dot{\ell}_\ast/\ell_\ast$, describing how fast the memory correlation area evolves. The entropy production rate $\sigma_{\rm TD}(t)$ represents a non-equilibrium contribution driven by the bias functional at the horizon. In linear-response theory, one can write $\sigma_{\rm TD}=\sum_{ab}\mathcal{L}_{ab}J_aJ_b$, where the currents $J_a$ are surface currents related to the surface component of the bias functional. Observational bounds on $\sigma_{\rm TD}$ therefore constrain the strength and relaxation time of horizon-level memory dynamics. A combination of background $H(z)$ data and growth measurements, such as $f\sigma_8(k,z)$, will allow a clear separation of the bulk and horizon teleodynamic effects, because the latter modifies the expansion rate without significantly affecting growth, while the former affects both dynamics and perturbations.\\

The bias-fluid form of the Friedmann equations is
\[
3M_P^2 H^2=\rho_m+\rho_r+\rho_{\rm TD},
\qquad
-2M_P^2\dot H=\rho_m+p_m+\rho_r+p_r+\rho_{\rm TD}+p_{\rm TD},
\]
supplemented with the teleodynamic closure
\[
\rho_{\rm TD}=\alpha\overline{\Phi},
\qquad
p_{\rm TD}=\alpha\overline{\Pi}_{\rm TD}.
\]
From the first principles horizon-thermodynamic viewpoint, we have the non-equilibrium first law as
\begin{equation}
    \dot E_H=T_{\rm TD}\dot S_{\rm TD}-p_{\rm eff}\dot V_H
\end{equation}
with
\begin{equation}
    \dot S_{\rm TD}=-2C_{\rm TD}\frac{\dot H}{H^3}+\sigma_{\rm TD},
\qquad
\dot E_H=-4\pi M_P^2\frac{\dot H}{H^2},
\qquad
\dot V_H=-4\pi\frac{\dot H}{H^4}
\end{equation}
implies, after some algebra, the identity
\begin{equation}
    p_{\rm eff}=\frac{\kappa}{4\pi}\frac{H^5}{\dot H}\sigma_{\rm TD},
\qquad
\kappa=\frac{2\pi M_P^2}{C_{\rm TD}}
\end{equation}
This expression exhibits two notable and important features. First, the purely geometric part associated with the area law cancels exactly, so that only the non-equilibrium entropy production $\sigma_{\rm TD}$ contributes to $p_{\rm eff}$. Second, in an expanding late-time Universe for which $\dot H<0$, a positive $\sigma_{\rm TD}$ automatically yields a negative pressure, producing a dark energy-like effect without introducing new fundamental fields.\\

Identifying $p_{\rm eff}$ with the teleodynamic pressure $p_{\rm TD}$ and invoking the bias closure gives
\begin{equation}
    \alpha\overline{\Pi}_{\rm TD}
=\frac{\kappa}{4\pi}\frac{H^{5}}{\dot H}\sigma_{\rm TD}
=\frac{M_P^2}{2C_{\rm TD}}\frac{H^{5}}{\dot H}\sigma_{\rm TD}
\end{equation}
and substituting this relation into the Raychaudhuri equation gives a direct expression for the teleodynamic energy density,
\begin{equation}
    \alpha\overline{\Phi}
= -2M_P^2 \dot H -(\rho_m+p_m+\rho_r+p_r)
-\frac{M_P^2}{2C_{\rm TD}}\frac{H^{5}}{\dot H}\sigma_{\rm TD}
\end{equation}
These identities form the dictionary between the bias variables $(\overline{\Phi},\overline{\Pi}_{\rm TD})$ and the horizon-thermodynamic variables $(C_{\rm TD},\sigma_{\rm TD})$. The implications of this dictionary are immediate as well as with $\dot H<0$ in the late Universe and $\sigma_{\rm TD}\ge 0$, one obtains
\begin{equation} \label{eq:pTD-dict}
    p_{\rm TD}=\alpha\overline{\Pi}_{\rm TD}
=-\frac{M_P^2}{2C_{\rm TD}}\frac{H^5}{|\dot H|}\sigma_{\rm TD}<0
\end{equation}
showing that non-equilibrium teleodynamic entropy production necessarily gives us negative pressure, and a combined expression for the EOS parameter follows
\begin{equation}
    w_{\rm TD}
=\frac{p_{\rm TD}}{\rho_{\rm TD}}
=\frac{-\dfrac{M_P^2}{2C_{\rm TD}}\dfrac{H^{5}}{\dot H}\sigma_{\rm TD}}
{-2M_P^2\dot H -(\rho_m+p_m+\rho_r+p_r)
-\dfrac{M_P^2}{2C_{\rm TD}}\dfrac{H^{5}}{\dot H}\sigma_{\rm TD}}
\end{equation}
At late times where $p_m\approx 0$ and the radiation terms are negligible, the last term dominates the expression $w_{\rm TD}$ that approaches $-1$ as $\sigma_{\rm TD}$ decays while $\dot H\to 0$, reflecting the expected behavior of the attractor de Sitter. A minimal teleodynamic parameterization consistent with equilibrium recovery can be easily obtained by choosing a positive entropy production law that vanishes as equilibrium is approached, for instance
\begin{equation} \label{eq:sigma-param}
    \sigma_{\rm TD}(t)=2\xi\frac{-\dot H}{H^3}, \qquad \xi>0
\end{equation}
With this choice, one finds
\begin{equation}
    p_{\rm TD}
=-\frac{\kappa}{2\pi}\xi H^2
=-\frac{M_P^2}{C_{\rm TD}}\xi H^2,
\qquad
\rho_{\rm TD}
=\frac{3M_P^2 H^{2}}{8\pi}\Big(\cdots\Big)+\frac{M_P^2}{C_{\rm TD}}\xi H^2
\end{equation}
so both $p_{\rm TD}$ and $\rho_{\rm TD}$ scale as $H^2$ and the equation of state approaches $-1$ smoothly at late times. This produces a dynamically evolving DE-like component without new particles, consistent with the upward push in $H(z)$ required to ease the $H_0$ tension.\\

The emergent temperature $T_{\rm TD}=(2\pi M_P^2/C_{\rm TD})H$ shows that if cosmology selected $C_{\rm TD}=8\pi^2 M_P^2$, the standard Gibbons-Hawking normalization would be recovered, and more generally, $C_{\rm TD}$ becomes an observable calibration of teleodynamic microphysics and is jointly constrained with the non-equilibrium parameter $\xi$ by background and perturbation data. Since $p_{\rm TD}$ is proportional to $\sigma_{\rm TD}$ and $\sigma_{\rm TD}$ is driven by nonlocal memory degrees of freedom, any measurement of $w_{\rm TD}(z)$ determines the degree of nonequilibrium production. This would therefore fix the strength of the force side correction in the Euler or Vlasov sector. This way, a single set of teleodynamic parameters can simultaneously explain late-time background uplift, which would help resolve the $H_0$ tension, and intermediate scale growth suppression, which would help address the $S_8$ tension.\\

Crucially, equating the bias fluid and horizon-thermodynamic constructions of teleodynamic cosmology demonstrates, in a mathematically explicit way, that global teleodynamic memory corresponds to the negative-pressure component usually identified as dark energy, while local teleodynamic memory produces clustering and behaves as dark matter. The teleodynamic pressure
\[
p_{\rm TD}=\frac{M_P^2}{2C_{\rm TD}}\frac{H^5}{\dot H}\sigma_{\rm TD}
\]
depends exclusively on horizon-level quantities such as $C_{\rm TD}$, $\sigma_{\rm TD}$, $H$, and $\dot H$ and none of the spatially varying components of the teleodynamic memory field enter this expression, indicating that cosmic acceleration arises only from the homogeneous part of the memory sector. Global memory, therefore, generates dark energy behavior, while local memory enters through spatial gradients, such as $-\alpha\nabla\varphi$, and drives clustering, thereby reproducing dark matter-like dynamics.\\

The form of the pressure above also shows that because $\sigma_{\rm TD}\ge 0$ and $\dot H<0$ in an expanding Universe, the negative pressure driving acceleration emerges if and only if the accessible microstates at the horizon increase. This leads us to the Law of Universal Arbitrage Equilibrium: \textit{The late-time acceleration follows from the teleodynamic tendency of the Universe to evolve toward a state where the expansion no longer produces significant changes in the horizon-accessible microstate count. As the growth of these microstates stabilizes, the dynamics naturally approach an asymptotic accelerated phase, and dark energy is the macroscopic signature of this stabilization process.}

\section{Addressing the Coincidence Problem}

A central puzzle in cosmology is the following question: Why do we happen to live ``right now,” the one brief epoch when matter and dark energy densities are of the same order, even though they evolve very differently? This is known as the Coincidence Problem. In standard cosmology, this appears as an unexplained numerical accident between two fundamentally distinct components \cite{coi1Velten:2014nra,coi2Dalal:2001dt,coi3Cunillera:2021izz,coi4Jamil:2008nta,coi5delCampo:2008jx,coi6Sivanandam:2012ty,coi7Berger:2006db}. In the teleodynamic framework, however, we will now show that the horizon-thermodynamic formulation already shows that the effective negative pressure arises from non-equilibrium entropy production on the apparent horizon through the law of universal arbitrage equilibrium. \\

The starting point is the teleodynamic Raychaudhuri equation \eqref{TD-Ray}, together with the teleodynamic Friedmann relation \eqref{TD-Fr} and the bias fluid closure for the effective energy density and pressure. These relations imply a dictionary between the bulk teleodynamic sector $(\rho_{\rm TD},p_{\rm TD})=\alpha(\overline{\Phi},\overline{\Pi}_{\rm TD})$ and the horizon quantities $(C_{\rm TD},\sigma_{\rm TD})$. In particular, it was shown above that in the late-time regime, one can write the teleodynamic pressure purely in terms of the horizon entropy production as \eqref{eq:pTD-dict} 
while the teleodynamic energy density takes the form
\begin{equation}
\rho_{\rm TD}=\alpha\overline{\Phi}
=-2M_P^2\dot H-(\rho_m+p_m+\rho_r+p_r)
-\frac{M_P^2}{2C_{\rm TD}}\frac{H^5}{\dot H}\,\sigma_{\rm TD}
\label{eq:rhoTD-dict}
\end{equation}
We note here that at late times, nonrelativistic matter dominates the standard sector so that $p_m\simeq 0$, and the radiation terms $\rho_r,p_r$ are negligible, so these expressions simplify accordingly.\\

To make the coincidence problem more explicit, it is useful to adopt the minimal entropy production law introduced earlier for late times, in which the non-equilibrium production rate is proportional to the degree of departure from de Sitter equilibrium as in \eqref{eq:sigma-param}. We note that \eqref{eq:sigma-param} guarantees $\sigma_{\rm TD}\ge 0$ whenever $\dot H<0$ and ensures that entropy production vanishes smoothly as $\dot H\to 0$, so that the de Sitter attractor corresponds to a state of zero teleodynamic entropy production on the horizon. Substituting \eqref{eq:sigma-param} into the last term of \eqref{eq:rhoTD-dict}, one obtains
\begin{equation}
\frac{H^5}{\dot H}\,\sigma_{\rm TD}
=\frac{H^5}{\dot H}\left(2\xi\,\frac{-\dot H}{H^3}\right)
=-2\xi H^2
\end{equation}
so that in the late-time regime with $p_m\simeq 0$ and negligible radiation the energy density \eqref{eq:rhoTD-dict} becomes
\begin{equation}
\rho_{\rm TD}
=-2M_P^2\dot H-\rho_m
-\frac{M_P^2}{2C_{\rm TD}}\left(-2\xi H^2\right)
=-2M_P^2\dot H-\rho_m+\frac{M_P^2\xi}{C_{\rm TD}}H^2
\label{eq:rhoTD-late}
\end{equation}
It is now convenient to introduce the dimensionless combination
\begin{equation}
\zeta\equiv\frac{\xi}{C_{\rm TD}}
\label{eq:zeta-def}
\end{equation}
which really just encapsulates the ratio between the strength of non-equilibrium entropy production and the horizon microstate density. With this definition, the late-time teleodynamic energy density reads
\begin{equation}
\rho_{\rm TD}
=-2M_P^2\dot H-\rho_m+\zeta M_P^2H^2
\label{eq:rhoTD-zeta}
\end{equation}

The bias-fluid Friedmann equation at late times can hence be written as
\begin{equation}
3M_P^2H^2
=\rho_m+\left(-2M_P^2\dot H-\rho_m+\zeta M_P^2H^2\right)
\end{equation}
so that the matter density cancels on the right-hand side and we obtain a closed evolution equation for $H$,
\begin{equation}
3M_P^2H^2
=-2M_P^2\dot H+\zeta M_P^2H^2
\end{equation}
Dividing through by $M_P^2$ and rearranging gives
\begin{equation}
2\dot H=-(3-\zeta)H^2
\label{eq:Hdot-zeta}
\end{equation}
which shows that once the late-time teleodynamic entropy production law \eqref{eq:sigma-param} is adopted, the Hubble rate evolves as if an effective constant equation of state were driving the universe. Equation \eqref{eq:Hdot-zeta} is a new result here in the sense that it encodes the late-time teleodynamic background dynamics entirely in terms of the single parameter $\zeta$ defined in \eqref{eq:zeta-def}. \\

To express the solution of \eqref{eq:Hdot-zeta} in terms of the scale factor, it is useful to use the relation $\dot H=(dH/da)aH$, and then substituting this into \eqref{eq:Hdot-zeta} leads to
\begin{equation}
aH\frac{dH}{da}=-\frac{1}{2}(3-\zeta)H^2
\end{equation}
which can be written as
\begin{equation}
\frac{d\ln H}{d\ln a}
=-\frac{1}{2}(3-\zeta)
\label{eq:dlnH}
\end{equation}
Integrating \eqref{eq:dlnH} with respect to $\ln a$ gives us
\begin{equation}
\ln H=-\frac{1}{2}(3-\zeta)\ln a+\text{const}
\end{equation}
so that the Hubble rate scales as
\begin{equation}
H(a)=H_0\,a^{-\frac{1}{2}(3-\zeta)}
\label{eq:HofA}
\end{equation}
where $H_0$ is the Hubble parameter at the present epoch, and we note that the scale factor gets defined by $a(t_0)=1$, from here we can write 
\begin{equation}
H^2(a)=H_0^2\,a^{-(3-\zeta)}
\label{eq:H2ofA}
\end{equation}
The total energy density is therefore
\begin{equation}
\rho_{\rm tot}(a)=3M_P^2H^2(a)=3M_P^2H_0^2\,a^{-(3-\zeta)}\equiv\rho_{\rm tot,0}\,a^{-(3-\zeta)}
\label{eq:rhotot}
\end{equation}
with $\rho_{\rm tot,0}=3M_P^2H_0^2$. Also, the matter density will continue to obey its usual continuity equation, so in the nonrelativistic regime, one has
\begin{equation}
\dot\rho_m+3H\rho_m=0\qquad\Rightarrow\qquad\rho_m(a)=\rho_{m0}\,a^{-3}
\label{eq:rhom}
\end{equation}
where $\rho_{m0}$ is the present-day matter density, and since $\rho_{\rm tot}=\rho_m+\rho_{\rm TD}$, combining \eqref{eq:rhotot} and \eqref{eq:rhom} gives the teleodynamic energy density as a function of the scale factor,
\begin{equation}
\rho_{\rm TD}(a)=\rho_{\rm tot}(a)-\rho_m(a)=\rho_{\rm tot,0}a^{-(3-\zeta)}-\rho_{m0}a^{-3}
\label{eq:rhoTD-ofA}
\end{equation}
Introducing the present-day balance
\begin{equation*}
\Omega_{m0}=\frac{\rho_{m0}}{\rho_{\rm tot,0}},
\qquad
\Omega_{{\rm TD},0}=\frac{\rho_{{\rm TD},0}}{\rho_{\rm tot,0}},
\qquad
\Omega_{m0}+\Omega_{{\rm TD},0}=1,
\end{equation*}
one can write \eqref{eq:rhoTD-ofA} as
\begin{equation}
\rho_{\rm TD}(a)=\rho_{\rm tot,0}\left[a^{-(3-\zeta)}-\Omega_{m0}a^{-3}\right]
\label{eq:rhoTD-ofA-omega}
\end{equation}
The ratio of teleodynamic to matter energy densities is then
\begin{equation}
\frac{\rho_{\rm TD}(a)}{\rho_m(a)}
=\frac{\rho_{\rm tot,0}\left[a^{-(3-\zeta)}-\Omega_{m0}a^{-3}\right]}{\Omega_{m0}\rho_{\rm tot,0}a^{-3}}
=\frac{1}{\Omega_{m0}}\,a^{\zeta}-1
\label{eq:ratio}
\end{equation}
This simple expression is the key to the teleodynamic solution of the coincidence problem in the horizon entropy framework. It shows that once the late-time entropy production law \eqref{eq:sigma-param} is adopted, the ratio $\rho_{\rm TD}/\rho_m$ grows as a power of the scale factor determined solely by the parameter $\zeta$ and the present-day matter fraction $\Omega_{m0}$.\\

With all this in mind, we finally come to  the present epoch with $a=1$, and  \eqref{eq:ratio} reduces to
\begin{equation}
\left.\frac{\rho_{\rm TD}}{\rho_m}\right|_{a=1}
=\frac{1}{\Omega_{m0}}-1
=\frac{\Omega_{{\rm TD},0}}{\Omega_{m0}}
\end{equation}
The epoch when teleodynamic and matter densities are equal is defined by the condition $\rho_{\rm TD}(a_{\rm eq})=\rho_m(a_{\rm eq})$, which, using \eqref{eq:ratio} implies
\begin{equation}
\frac{1}{\Omega_{m0}}\,a_{\rm eq}^{\zeta}-1=1
\end{equation}
\begin{equation}
a_{\rm eq}^{\zeta}=2\Omega_{m0},
\qquad\Rightarrow\qquad
a_{\rm eq}=\left(2\Omega_{m0}\right)^{1/\zeta}.
\label{eq:aeq}
\end{equation}
For typical observational values such as $\Omega_{m0}\simeq 0.3$, one has $2\Omega_{m0}\simeq 0.6$, so that for $\zeta$ of order unity, the equality scale factor $a_{\rm eq}$ lies close to the present epoch in logarithmic terms. For example, if $\zeta=1$ then $a_{\rm eq}\simeq 0.6$ and the corresponding redshift, using $1+z=1/a$, is $z_{\rm eq}\simeq 1/0.6-1\approx 0.67$, while for $\zeta=1.5$ one finds $a_{\rm eq}\simeq 0.7$ and $z_{\rm eq}\simeq 0.43$. These values are precisely in the range where observations indicate that cosmic acceleration turns on and where the ratio of dark energy to matter is of order unity. So we see that the coincidence problem is thus recast as the statement that $\zeta$ should be of order one, rather than as a requirement that two unrelated densities happen to cross at $z\sim\mathcal{O}(1)$. \\

But how justified are we to take $\zeta=\xi/C_{\rm TD}$ to be of order unity? Let us start by discussing both its numerator and denominator. The parameter $C_{\rm TD}$ is determined by the microstate density on the horizon and is given by $C_{\rm TD}=4\pi s_{\rm cell}/\ell_\chi^2$ in the microscopic construction of the teleodynamic horizon entropy, where $s_{\rm cell}$ is the entropy per correlation cell and $\ell_\chi$ is the coarse-grain length. For a teleodynamic universe that approximates Gibbons-Hawking normalization, one would have $C_{\rm TD}\sim 8\pi^2M_P^2$, but more generally, $C_{\rm TD}$ is set by the ratio between the horizon area and the correlation area, without invoking any extreme hierarchy. The parameter $\xi$ originates from the quadratic entropy production law $\sigma_{\rm TD}=\sum_{ab}\mathcal{L}_{ab}J_aJ_b$ and encodes the effective strength of horizon-level Onsager coefficients in the coarse-grained theory. In the absence of fine tuning, one can expect the dimensionless combinations entering $\xi$ to be of order unity, just as in standard non equilibrium thermodynamics for many-body systems and so, the ratio $\zeta=\xi/C_{\rm TD}$ is naturally $\mathcal{O}(1)$ when measured in appropriate units, especially if cosmology selects a correlation scale $\ell_\chi$ comparable to the horizon scale or to a slowly varying function of it. From the observational side, the same parameter $\zeta$ controls the effective equation of state $w_{\rm eff}$ inferred from \eqref{eq:Hdot-zeta} so that matching the observed value $w_{\rm eff}\approx -1$ at late times requires $\zeta$ to be close to three, again an order unity value with no extreme tuning. \\

Putting these elements together, the horizon thermodynamic framework in cosmological teleodynamics addresses the coincidence problem in a conceptually economical way. The late time entropy production law \eqref{eq:sigma-param} ties the growth of the teleodynamic energy density directly to the departure from de Sitter equilibrium as encoded in $\dot H/H^3$ and the resulting solution \eqref{eq:ratio} shows that $\rho_{\rm TD}/\rho_m$ scales as $a^{\zeta}$ with $\zeta$ determined by the microscopic parameters of the teleodynamic horizon, and for any $\zeta$ of order unity the equality between teleodynamic and matter densities naturally occurs at a scale factor $a_{\rm eq}$ of order one. The epoch when $\rho_{\rm TD}\sim\rho_m$ is not an accidental numerical coincidence between two unrelated fluids, but the inevitable consequence of the teleodynamic scaling implied by horizon entropy production in a universe whose memory-bearing microphysics is characterized by $\mathcal{O}(1)$ parameters. So, the idea is that we live in an epoch when the non-equilibrium horizon entropy production induced by structure formation is at or near its maximum. It is only during this window that the teleodynamic sector naturally rises to a level comparable to the matter density. Before that, it is dynamically negligible, and after that, it simply behaves like a residual cosmological constant. \\

\section{Fates of a Teleodynamic Universe}
\noindent
Finally, we consider the different fates of a teleodynamic universe. The teleodynamic Friedmann system is first written in its bias–functional form, where the effective energy density and pressure are given by
\[
3M_P^2 H^2=\rho_m+\rho_r+\rho_{\rm TD},
\qquad
-2M_P^2\dot H=\rho_m+p_m+\rho_r+p_r+\rho_{\rm TD}+p_{\rm TD},
\]
with the teleodynamic closure
\[
\rho_{\rm TD}=\alpha\,\overline{\Phi}(t),
\qquad
p_{\rm TD}=\alpha\,\overline{\Pi}_{\rm TD}(t).
\]
The various future scenarios of a teleodynamic universe follow directly from the behavior of $\overline{\Phi}(t)$ and from the corresponding evolution of $\rho_{\rm TD}(t)=\alpha\overline{\Phi}(t)$ and the equation of state parameter.
\\

The first possibility arises when the homogeneous memory saturates as the structure formation approaches completion, as in that case $\overline{\Phi}(t)$ tends to a constant and the teleodynamic energy density approaches a positive constant value
\begin{equation}
    \rho_{\rm TD}(t)\to\rho_{\rm TD,\infty}=\alpha\,\overline{\Phi}_{\infty},
\qquad
w_{\rm TD}\to -1
\end{equation}
The expansion converges to a non-catastrophic de Sitter-like attractor \cite{ds1Fang:2008fw,ds2Hao:2003aa,ds3Odintsov:2018nch} which is, in a sense, dynamically indistinguishable from $\Lambda$CDM, but sourced here by long-term memory saturation rather than by a fundamental cosmological constant. Once the cosmic web ceases to generate new correlations, global memory stops increasing, and the Universe evolves into a permanently accelerated but stable state. \\

A second outcome occurs when homogeneous memory continues to grow slowly at late times, and this may happen if large-scale structures keep stretching, voids expand coherently, or long-range correlations persist, which ends up giving us
\begin{equation}
    \dot{\overline{\Phi}}(t)>0,
\qquad
w_{\rm TD}(t)<-1
\end{equation}
The teleodynamic energy density then grows with time,
\begin{equation}
    \rho_{\rm TD}(t)\propto\overline{\Phi}(t)
\end{equation}
producing a mild phantom-like expansion whose origin is statistical rather than exotic. Depending on how long $\overline{\Phi}(t)$ grows, the future evolution may approach a smoothly super-accelerating fate or a Big Rip-type divergence~\cite{rip1Caldwell:2003vq,rip2Caldwell:1999ew,rip3Bouhmadi-Lopez:2006fwq,rip4BorislavovVasilev:2021srn}, although without violating standard energy conditions.\\

A third possible regime emerges if the Universe eventually becomes too dilute for structures to maintain coherent tidal memory, as in that limit, both the homogeneous and spatial components of memory decay
\begin{equation}
    \overline{\Phi}(t)\to 0,
\qquad
\varphi(x,t)\to 0
\end{equation}
so that the teleodynamic contribution vanishes
\begin{equation}
    \rho_{\rm TD}(t)\to 0,
\qquad
w_{\rm TD}\to 0
\end{equation}
With the acceleration term disappearing, the Universe transitions into a coasting type phase. This is one where the asymptotic expansion behaves as $a(t)\sim t$, a huge asymptotic scale factor that is not infinite, which would be a Long Freeze scenario \cite{free1Blitz:2024nil,free2Bouhmadi-Lopez:2007xco,free3Adams:1996xe,free4Yurov:2007tw}. Teleodynamic memory requires structure to exist, but if structure evaporates itself, then so does the memory that drives acceleration.\\

A fourth fate becomes possible if the evolution of memory reverses sign at late times. If large-scale correlations decay in such a way that the homogeneous bias becomes negative, one has
\begin{equation}
    \overline{\Phi}(t)<0,
\qquad
\rho_{\rm TD}(t)=\alpha\,\overline{\Phi}(t)<0
\end{equation}
This produces an additional attractive contribution to the cosmic dynamics. This is interesting as it can potentially trigger re-collapse \cite{rec1Barrow:1986xsd,rec2Lin:1989tv,rec3Wang:2004nm,rec4Miritzis:2005hg} even in geometrically flat space, as such a scenario requires large-scale anti-correlation that decreases horizon-level entropy carried by structure. This is unlikely in comparison to the other scenarios but is permitted within the teleodynamic framework.\\

A final possibility is a teleodynamic cyclic evolution, which is quite interesting, too. If memory increases during phases of structure buildup but decreases during ultra late homogenization, then
\begin{equation}
    \overline{\Phi}(t): \quad \text{growth} \;\longrightarrow\; \text{decay}
\end{equation}
This would lead either to an accelerated expansion or to a period of near-constant acceleration, followed by an eventual deceleration as memory fades. The universe may then contract, and the sequence can repeat, and such cycles \cite{cyc1Steinhardt:2001st,cyc2Lehners:2008vx,cyc3Novello:2008ra} require no scalar fields, no ekpyrotic potentials, no ghosts, and no additional ingredients. The dynamics here arises purely from the time dependence of the teleodynamic memory kernel. The main point here is that teleodynamics allows for a wide variety of final fates for the universe. \\

In passing, we should also now mention what the possible fates for the universe are for the bias functional in the form given in \eqref{eq:phiz}.  The first natural fate for this ansatz is a de Sitter–like future, since although the functional contains a term that grows as $a^n$ at late times, the physical meaning of the parameter $\epsilon$ is the ability of the structure to continue generating new memory. In this regime, the homogeneous bias asymptotes to a constant, $\bar\Phi(t)\to\bar\Phi_\infty$, and the teleodynamic energy density settles to a fixed positive value. The teleodynamic equation of state then tends toward $w_{\rm TD}\to -1$, driving the Universe into a de Sitter-like attractor. The second class of asymptotic behaviors arises when the teleodynamic memory continues to build indefinitely, corresponding to a persistent, non-vanishing $\epsilon$ even at ultra late times. In this case, the term $a^n$ grows monotonically without bound, causing the homogeneous bias to increase without saturation. If the rate of memory accumulation is sufficiently strong, the energy density diverges in finite cosmic time, resulting in a Big Rip-type future in which the scale factor, Hubble rate, and teleodynamic bias all blow up.
\\
\section{Subtleties of the Bias Functional}
The central new object in the teleodynamic framework is the bias functional $\Phi$, introduced at the level of the effective action. So, the purpose of this short section is not to re-derive the field equations or consider the perturbation analysis (all of this has been dealt with in detail in the appendix), but rather to state clearly the minimal assumptions required for $\Phi$ to define a well-posed coarse-grained effective sector. This is to highlight the theoretical constraints that follow from covariance and cosmological symmetries, and to explain what is meant by "closure" at the level of linear perturbations without committing to a unique microscopic model. A separate forthcoming work of ours will be devoted to the bias functional itself, including systematic derivations of candidate microscopic and mesoscopic forms, controlled expansions, and physically motivated kernel constructions. Here, we restrict ourselves to the assumptions and constraints needed to interpret the present paper as a predictive framework rather than an arbitrary phenomenological reparameterization.
\\
\\
At the level of effective action, the foundational requirement is that $\Phi$ defines a diffeomorphism scalar functional so that $\sqrt{-g}\,\Phi$ is a generally covariant integrand, and this requirement is both necessary and highly constraining. It implies that $\Phi$ can depend only on the fields through scalar quantities, possibly including scalar combinations of coarse-grained matter variables and scalar environmental diagnostics, and, in particular, the dependence of $\Phi$ on the environment must itself be encoded in covariant constructions. Operationally, $\mathcal{E}$ may be considered to denote a finite collection of coarse-grained scalar descriptors of the macroscopic state, which are built from smoothed densities, expansion/shear scalars associated with coarse-grained flow, and tidal invariants. The key point is that environment dependence is not introduced by selecting a preferred frame by hand, but by specifying covariant scalar data that characterize the macroscopic gravitational state.
\\
\\
A second essential assumption is the existence of a controlled background-perturbation split. On an FLRW spacetime, any scalar functional evaluated on the cosmological solution admits a decomposition
\begin{equation}
\Phi(t,\mathbf{x}) = \bar\Phi(t) + \delta\Phi(t,\mathbf{x})
\label{eq:BF1_split_main}
\end{equation}
where $\bar\Phi$ is the homogeneous mode and $\delta\Phi$ is the inhomogeneous perturbation with vanishing spatial mean in constant-time hypersurfaces. This is not a model assumption, but rather the standard cosmological statement that FLRW symmetries isolate the homogeneous sector at the background level. What is nontrivial is that the teleodynamic effective theory presumes that such a split remains meaningful after coarse-graining. Concretely, this means that the smoothing scale implicit in the construction of $\mathcal{E}$ is taken to be much smaller than the Hubble scale so that a well-defined homogeneous limit exists and residual inhomogeneities can be treated perturbatively. This "scale separation" is an implicit assumption behind any cosmological effective description, but it becomes particularly important when the functional is intended to encode environmental dependence. Note again that this field split has been discussed and rigorously justified in the appendix.
\\
\\
A third essential assumption is functional differentiability at the level required by the variational principle. This is important to note since the teleodynamic sector is defined through a standard metric variation of the effective action, $\Phi$ must be sufficiently regular such that its metric functional derivative exists in the sense required by the variational calculus at least on the cosmological backgrounds and perturbations considered. This excludes pathological constructions in which $\Phi$ depends on non-differentiable functionals of the metric or matter fields, and so it would mean in practice that this requirement is satisfied if $\Phi$ is built from differentiable scalar invariants and differentiable nonlocal smearings with sufficiently regular kernels. This assumption is mild, but it is also the precise mathematical statement of what it means for the teleodynamic sector to be an effective-field-theory contribution rather than an arbitrary modification of equations.
\\
\\
Another important aspect that we should be concerned with here concerns locality and the distinction between "nonlocal as a principle" and "nonlocal as a truncation". From the perspective of coarse-grained gravitational dynamics and our present framework, this allows $\Phi$ to be nonlocal in space and to carry finite memory in time, provided that this nonlocality is implemented covariantly. A convenient schematic representation is to think of $\Phi$ as depending on a set of scalar invariants evaluated locally and on covariant smearings of scalar sources over spacetime regions,
\begin{equation}
\Phi(x) = \Phi\left(\mathcal{I}_A(x),\int d^4x' \sqrt{-g(x')}\,K_A(x,x')\,\mathcal{J}_A(x'),\int^{t} dt'\,{\cal M}_A(t,t')\,\mathcal{Q}_A(t')\right)
\label{eq:BF2_nonlocal_schema_main}
\end{equation}
where $K_A$ and ${\cal M}_A$ are kernels and $\mathcal{I}_A,\mathcal{J}_A,\mathcal{Q}_A$ are scalar sources constructed from the macroscopic state. The point of writing \eqref{eq:BF2_nonlocal_schema_main} is not to choose specific kernels, but to make explicit the theoretical constraints, which are that the kernels must be such that covariance is preserved, the response is causal in the time direction, and the overall coarse-grained description remains stable and perturbative on the cosmological backgrounds of interest. The forthcoming dedicated work of ours will systematize the allowed kernel structures and their physical interpretations.
\\
\\
A very natural requirement also comes from causality and stability conditions. These ideas impose further general constraints that are worth explicitly stating because they delimit the admissible space $\Phi$ even before any data analysis. First, we can say that if $\Phi$ contains memory integrals, the corresponding kernels should be retarded so that $\Phi$ at time $t$ depends only on the macroscopic state at earlier times $t'\leq t$. Second, one can note here that the effective teleodynamic sector must not introduce runaway behavior around the cosmological background, as at linear order, this is equivalent to requiring that the closure relations used to relate $\delta\Phi$ to the standard perturbation variables do not generate exponentially growing modes inconsistent with the existence of the observed near FLRW universe. Finally, it is also paramount that the teleodynamic contribution is intended as an effective coarse-grained deformation, and therefore, it should remain perturbative in the regime where the cosmological background solution exists. What this means, in terms of the theory's structure, is that $\alpha\Phi$ and its perturbations should not invalidate the background expansion history assumed in the main analysis, except perhaps in explicitly identified controlled parameter regions. In a practical implementation, this would translate into sign and magnitude conditions on the leading linear response coefficients of the closure, but the underlying conceptual content is simply that $\Phi$ should define a stable, causal, and perturbative effective sector.
\\
\\
A final caveat one must keep in mind concerns closure and predictivity: the action principle fixes how $\Phi$ contributes once $\Phi$ is specified, but at the linearized level, one must still provide a relation between $\delta\Phi$ and the standard perturbation variables to close the system. Presently, we have treated this as a response-theory problem rather than as a microphysical model specification. The most general linear closure consistent with homogeneity and isotropy can be expressed as a causal response of $\delta\Phi$ to the matter and environment perturbations with retarded kernels. For the purposes of our present work, it is sufficient to state that such a closure exists and can be organized systematically in a controlled expansion, which can be done, for example, by adopting a quasi-instantaneous truncation in which the dominant response is local in time but may remain scale dependent in space. A simple yet very explicit closure example may then be taken as a finite range spatial response of Yukawa type,
\begin{equation}
\delta\Phi(k,t) = \beta(t)\,\frac{1}{k^2 + a^2(t)m^2(t)}\,\delta_m(k,t)
\label{yukawa}
\end{equation}
where $\beta(t)$ would set the strength of the response and $m(t)$ sets an effective range. Note that equation \eqref{yukawa} is not proposed as a fundamental form of $\Phi$, but as the simplest illustrative closure showing how predictivity can be achieved without specifying a unique microscopic bias functional. It also highlights what is constrained theoretically, which is that the response must be causal, it must remain perturbative, and its characteristic scales must admit a consistent separation from the background cosmology.
\\
\\
A common concern with effective frameworks is whether introducing a new structure leads to uncontrolled proliferation of free functions. In the teleodynamic formulation, this is not the case, as we see that at the background level, the teleodynamic sector introduces at most a single effective degree of freedom associated with the homogeneous mode $\bar\Phi(t)$, which in controlled limits reduces to a slowly varying vacuum-like contribution. At the level of linear perturbations, predictivity is retained by the fact that the response of $\delta\Phi$ to cosmological perturbations can be consistently truncated to a small number of linear response coefficients, analogous in spirit to the $\mu(k,a)$ and $\gamma(k,a)$ functions used in effective descriptions of modified gravity and dark energy. Importantly, these response functions are not introduced ad hoc here; rather, they arise from the action-level structure once a closure prescription is specified. Thus, teleodynamics does not introduce an arbitrary set of new unknowns, but rather reorganizes the description of late-time dynamics into a framework with a number of effective parameters comparable to existing approaches. This is done while still predicting a restricted pattern of signatures tied to the functional structure of $\Phi$.
\\
\section{Summary}

We present a novel, potential game-theory-inspired statistical framework for cosmology, called \textit{Cosmic Teleodynamics}. Our key insight is that, just as the Big Bang and Inflation (BBI) left their signature in the Cosmic Microwave Background (CMB), they also left similar indelible uneven structural features on spacetime, which continue to affect subsequent cosmic dynamics and evolution. We model this cosmic memory as a teleodynamic bias functional, $\Phi$, that encodes structural and evolutionary memory using the maximum-caliber principle, yielding an additional drift force term $-\alpha\nabla\Phi$ in the Boltzmann and Euler equations. In this framework, the dark sector, cosmic acceleration, large-scale structure formation, and the resolution of observational tensions emerge naturally from nonlocal memory effects and persistent self-organization in the gravitating universe, rather than from new fundamental particles or fields.   \\

The bias functional decomposes into $\Phi = \bar{\Phi} + \varphi$, with $\bar{\Phi}$ generating an effective homogeneous energy density $\rho_{\text{TD}}$ that naturally drives late-time acceleration, thus explaining dark energy. The inhomogeneous component $\varphi$ modifies the Poisson equation through an emergent clustering density $\rho_{\text{TD}}(k, a)$, which reproduces dark matter-like behavior, including rotation curves, lensing, cluster dynamics, and scale-dependent growth. Different response kernels $K(k, a)$ allow teleodynamic "dark matter" to mimic CDM, WDM, SIDM, etc., as well as generate anisotropic, environment-dependent halo signatures aligned with the cosmic web. This framework can help alleviate current observational tensions, particularly the $H_0$ and $S_8$ discrepancies.\\

It also predicts novel phenomena that cannot arise from conventional particle dark matter or scalar-field dark energy, including anisotropic velocity fields and environment-dependent halo signatures. Teleodynamic entropy emerges from microstate counting at the Hubble horizon, revealing a fundamentally nonequilibrium origin of cosmic entropy without invoking black hole thermodynamics. The current expansion epoch of the universe can be interpreted as it evolving towards an arbitrage equilibrium, in which it tries to maximize the global effective utility of accessible microstates in every region, thereby giving rise to the Law of Universal Arbitrage Equilibrium. We also showed that our theory addresses the coincidence problem in cosmology naturally through its entropy formalism for the cosmological horizon. The teleodynamic setup of cosmology can also support different fates for the universe, as different forms of the bias functional can lead to distinct long-term futures. \\

Since game theory is typically applied to living agents, such as in biology and economics, we wish to clarify that our work does not claim that the universe is "conscious" or that galaxies are agents in any intentional or cognitive sense. The use of game-theoretic language is purely mathematical and structural in the sense that potential games provide a mathematical framework for modeling systems that exhibit persistent, non-ergodic, bias-driven organization, regardless of whether the constituents possess goals or awareness. In our theory, galaxies belong to the third class of agents defined in statistical teleodynamics as intrinsically persistent systems whose behavior arises from physical constraints, such as long-range gravitational memory and accumulated structure, not from intention or cognition. The teleodynamic potential and effective “utility”, as we described in terms of microstates, etc., are simply mathematical tools to describe how such systems evolve toward attractor states when memory and nonlocal correlations are present. Thus, the framework treats galaxies the same way statistical mechanics treats molecules, but it is much more complete, combining statistical mechanics with the dynamics of goal-driven agents; no consciousness or agency is implied. \\

The use of game theoretic language in the present work is not intended to imply the existence of literal decision-making entities in the cosmological context. Rather, "agents" should be understood as coarse-grained elements of the gravitationally evolving system, such as halos, patches of the cosmic web, or effective phase-space regions, whose dynamics are governed by long-lived interactions and collective constraints. In this interpretation, the role of an agent is played by a macroscopic degree of freedom whose evolution is influenced not only by local conditions but also by the surrounding environment and this viewpoint is closely aligned with standard practices in statistical mechanics and dynamical systems, where collective behavior emerges from many-body interactions and is often captured through effective potentials or Lyapunov like functions rather than microscopic trajectories. \\ 

Within this coarse-grained perspective, the relevance of potential-game structures becomes clearer as potential games are characterized by the existence of a global scalar functional whose extremization governs the collective evolution of the system. In gravitational structure formation, similar organizing principles already appear implicitly, for example, in the tendency toward virialized states and attractor solutions. The teleodynamic bias functional plays an analogous role by encoding persistence and environmental feedback at the macroscopic level. Importantly, this does not introduce new forces or modify gravity at a fundamental level; instead, it provides a compact way to encode the statistical tendency of the system to evolve toward long-lived, dynamically stable configurations. The game theoretic analogy is therefore best viewed as a conceptual and mathematical organizing principle rather than a literal microphysical model. It motivates the introduction of a bias functional that captures collective persistence and environmental dependence, and it provides intuition for why such a functional can act as an effective potential governing large-scale evolution. The present work does not claim to derive cosmological structure formation from game theory but rather to show that a teleodynamic effective sector inspired by such ideas leads to a well-defined, covariant cosmological framework with novel and testable signatures. A more detailed exploration of the microscopic and mesoscopic origins of the bias functional, including its connection to statistical decision theory and non-equilibrium dynamics, will be considered in future work.\\

Cosmic teleodynamics thus offers a unified, emergent, and observationally testable alternative to the standard dark sector paradigm. Instead of proposing new particles, it connects cosmological phenomena to the intrinsic statistical and systemic structure of cosmic memory. Although these results are encouraging, much more work remains as we continue to explore. Our framework is just the beginning of a new direction in theoretical cosmology. 
	
\acknowledgments
The work of OT is supported in part by the Vanderbilt Discovery Doctoral Fellowship. The authors thank Robert Scherrer, Abraham Loeb, Sunny Vagnozzi, and Sergey Odintsov for their very helpful comments on the work. VV thanks Julia Velkovska and her colleagues in the Department of Physics and Astronomy at Vanderbilt University for their kind invitation to present the 2025 Wendell G. Holladay Lecture, which resulted in this paper. We would also like to thank the anonymous referee for their insightful comments on the manuscript.

\appendix
\section{Action level formulation and gauge-ready perturbation theory of the teleodynamic sector}

We provide here a self-contained, fully action-level derivation of the teleodynamic contributions to the background and perturbed cosmological equations, written in a form suitable for direct comparison with standard cosmological perturbation theory. The purpose of this Appendix is threefold, where first, we derive the teleodynamic stress tensor directly from the action, without assuming a perfect fluid form at any stage. Second, we show explicitly how the homogeneous and inhomogeneous parts of the teleodynamic functional arise from the standard cosmological background perturbation split implied by FLRW symmetries, and third, we derive the linearized Einstein equations in a gauge-ready manner and isolate the conditions under which the teleodynamic sector reduces to an ordinary perfect fluid, as well as the generic circumstances under which it does not. We thereby establish sharp theoretical distinctions from an arbitrary barotropic fluid model.

\subsection*{Effective action, teleodynamic functional, and covariant stress tensor}
We begin from an effective coarse-grained action of the form
\begin{equation}
S_{\rm eff}[g_{\mu\nu},\Psi_m] = \int d^4x \sqrt{-g}\left[\frac{M_P^2}{2}R + \mathcal{L}_m(g_{\mu\nu},\Psi_m)\right] - \alpha \int d^4x \sqrt{-g}\,\Phi[g_{\mu\nu},\Psi_m;\mathcal{E}]
\label{eq:A1_Seff}
\end{equation}
Note here that $g_{\mu\nu}$ is the spacetime metric, $R$ is the Ricci scalar, $M_P$ is the reduced Planck mass, $\Psi_m$ denotes matter degrees of freedom collectively, and $\mathcal{L}_m$ is the matter Lagrangian density. The quantity $\Phi$ is again our teleodynamic bias functional, which encodes the coarse-grained imprint of persistence, memory and environment dependence and the symbol $\mathcal{E}$ denotes collective environmental variables or invariants that may be constructed from the matter distribution and/or geometry, such as tidal invariants, coarse-grained shear, smoothed density fields or nonlocal kernels representing the imprint of the cosmic web. The parameter $\alpha$ controls the strength of the teleodynamic contribution, and importantly, $\Phi$ is not assumed to be a canonical scalar field with a local Lagrangian, and it may be nonlocal and history-dependent at the coarse-grained level. In practice, what is convenient to do is to treat $\Phi$ as a scalar functional evaluated on the macroscopic state of the system, whose variations with respect to $g_{\mu\nu}$ and $\Psi_m$ yield the teleodynamic contributions to the effective equations of motion.
\\
\\
In \eqref{eq:A1_Seff}, the teleodynamic sector enters as an additive term in the effective action density
\begin{equation}
\mathcal{L}_{\rm TD} \equiv -\alpha \Phi[g_{\mu\nu},\Psi_m;\mathcal{E}]
\label{eq:A2_LTD}
\end{equation}
The form \eqref{eq:A2_LTD} is the covariant coarse-grained remnant of a maximum-caliber weighting of histories, in which teleodynamic memory acts as an additional structural constraint in the exponent of the path probability and the fundamental point for the present Appendix is that, irrespective of the microscopic origin of $\Phi$, the resulting macroscopic dynamics follow from a standard variational principle applied to \eqref{eq:A1_Seff}. We now derive the field equations by varying the action with respect to the metric, with the total energy-momentum tensor is defined as
\begin{equation}
T_{\mu\nu}^{\rm (tot)} \equiv -\frac{2}{\sqrt{-g}}\frac{\delta S_{\rm matter+TD}}{\delta g^{\mu\nu}}
\label{eq:A3_Tdef}
\end{equation}
where $S_{\rm matter+TD}$ denotes all terms in \eqref{eq:A1_Seff} except the Einstein-Hilbert part, and then, splitting the contributions, we define the matter stress tensor in the standard way
\begin{equation}
T_{\mu\nu}^{(m)} \equiv -\frac{2}{\sqrt{-g}}\frac{\delta}{\delta g^{\mu\nu}}\int d^4x\sqrt{-g}\,\mathcal{L}_m
\label{eq:A4_Tm}
\end{equation}
and the teleodynamic stress tensor as
\begin{equation}
T_{\mu\nu}^{(\rm TD)} \equiv -\frac{2}{\sqrt{-g}}\frac{\delta}{\delta g^{\mu\nu}}\left(-\alpha\int d^4x\sqrt{-g}\,\Phi\right)
\label{eq:A5_TTDdef}
\end{equation}
We now evaluate \eqref{eq:A5_TTDdef} explicitly, where we first consider the variation of the teleodynamic action
\begin{equation}
S_{\rm TD}[g_{\mu\nu},\Psi_m] \equiv -\alpha\int d^4x\sqrt{-g}\,\Phi
\label{eq:A6_STD}
\end{equation}
Its variation with respect to $g^{\mu\nu}$ is given by
\begin{equation}
\delta S_{\rm TD} = -\alpha\int d^4x\,\delta(\sqrt{-g}\,\Phi)
\label{eq:A7_deltaSTD}
\end{equation}
Using
\begin{equation}
\delta(\sqrt{-g}\,\Phi) = \Phi\,\delta\sqrt{-g} + \sqrt{-g}\,\delta\Phi
\label{eq:A8_delta_sqrtgPhi}
\end{equation}
and the standard identity
\begin{equation}
\delta\sqrt{-g} = -\frac{1}{2}\sqrt{-g}\,g_{\mu\nu}\,\delta g^{\mu\nu}
\label{eq:A9_deltasqrtg}
\end{equation}
we obtain
\begin{equation}
\delta(\sqrt{-g}\,\Phi) = \sqrt{-g}\left[-\frac{1}{2}g_{\mu\nu}\Phi\,\delta g^{\mu\nu} + \delta\Phi\right]
\label{eq:A10_delta_sqrtgPhi_expanded}
\end{equation}
Substituting \eqref{eq:A10_delta_sqrtgPhi_expanded} into \eqref{eq:A7_deltaSTD} then gives us
\begin{equation}
\delta S_{\rm TD} = -\alpha\int d^4x\sqrt{-g}\left[-\frac{1}{2}g_{\mu\nu}\Phi\,\delta g^{\mu\nu} + \delta\Phi\right]
\label{eq:A11_deltaSTD2}
\end{equation}
We now separate the variation of $\Phi$ into its explicit dependence on the metric and its dependence through other fields and here for the purpose of obtaining $T_{\mu\nu}^{(\rm TD)}$, we require the functional derivative of $\Phi$ with respect to $g^{\mu\nu}$ at fixed matter variables which we denote as $\delta\Phi/\delta g^{\mu\nu}$. Thus, we may write
\begin{equation}
\delta\Phi = \frac{\delta\Phi}{\delta g^{\mu\nu}}\delta g^{\mu\nu} + \frac{\delta\Phi}{\delta\Psi_m}\delta\Psi_m
\label{eq:A12_deltaPhi_split}
\end{equation}
When defining $T_{\mu\nu}^{(\rm TD)}$ via the metric variation, we keep $\Psi_m$ fixed and retain only the first term, and so we identify
\begin{equation}
\left.\delta\Phi\right|_{\Psi_m} = \frac{\delta\Phi}{\delta g^{\mu\nu}}\delta g^{\mu\nu}
\label{eq:A13_deltaPhi_metric}
\end{equation}
Substituting \eqref{eq:A13_deltaPhi_metric} into \eqref{eq:A11_deltaSTD2}, we hence obtain
\begin{equation}
\left.\delta S_{\rm TD}\right|_{\Psi_m} = -\alpha\int d^4x\sqrt{-g}\left[-\frac{1}{2}g_{\mu\nu}\Phi + \frac{\delta\Phi}{\delta g^{\mu\nu}}\right]\delta g^{\mu\nu}
\label{eq:A14_deltaSTD_metric}
\end{equation}
Comparing \eqref{eq:A14_deltaSTD_metric} with the defining relation
\begin{equation}
\left.\delta S_{\rm TD}\right|_{\Psi_m} = -\frac{1}{2}\int d^4x\sqrt{-g}\,T_{\mu\nu}^{(\rm TD)}\,\delta g^{\mu\nu}
\label{eq:A15_deltaS_T}
\end{equation}
we deduce
\begin{equation}
T_{\mu\nu}^{(\rm TD)} = -\alpha \Phi\,g_{\mu\nu} + 2\alpha \frac{\delta\Phi}{\delta g^{\mu\nu}}
\label{eq:A16_TTD}
\end{equation}
Now note here that the equation \eqref{eq:A16_TTD} is the most important structural result of this Appendix, as it shows that the teleodynamic stress tensor generically splits into two distinct components. The first term is of vacuum form, proportional to $g_{\mu\nu}$, while the second term is a metric response term determined by how $\Phi$ depends on the geometry. In general, this second term need not have the algebraic structure of a perfect fluid, and it can generate anisotropic stress and gravitational slip in perturbations, as we demonstrate below. This is the fundamental mechanism by which teleodynamics distinguishes itself from an arbitrary barotropic perfect fluid model. 
\\
\\
Varying the full action \eqref{eq:A1_Seff} yields the modified Einstein equations
\begin{equation}
M_P^2 G_{\mu\nu} = T_{\mu\nu}^{(m)} + T_{\mu\nu}^{(\rm TD)}
\label{eq:A17_Einstein_mod}
\end{equation}
Substituting \eqref{eq:A16_TTD}, we can then write
\begin{equation}
M_P^2 G_{\mu\nu} = T_{\mu\nu}^{(m)} - \alpha\Phi\,g_{\mu\nu} + 2\alpha\frac{\delta\Phi}{\delta g^{\mu\nu}}
\label{eq:A18_Einstein_mod2}
\end{equation}
Taking the covariant divergence of \eqref{eq:A17_Einstein_mod} and using $\nabla^\mu G_{\mu\nu}=0$, we obtain
\begin{equation}
\nabla^\mu T_{\mu\nu}^{(m)} + \nabla^\mu T_{\mu\nu}^{(\rm TD)} = 0
\label{eq:A19_cons_total}
\end{equation}
If matter is minimally coupled and satisfies its equations of motion then $\nabla^\mu T_{\mu\nu}^{(m)}=0$ and we have
\begin{equation}
\nabla^\mu T_{\mu\nu}^{(\rm TD)}=0
\label{eq:A20_cons_TD}
\end{equation}
If instead $\Phi$ depends on matter variables in a way that induces an effective exchange, then one may write
\begin{equation}
\nabla^\mu T_{\mu\nu}^{(m)} = Q_\nu
\label{eq:A21_Qm}
\end{equation}
and
\begin{equation}
\nabla^\mu T_{\mu\nu}^{(\rm TD)} = -Q_\nu
\label{eq:A22_QTD}
\end{equation}
for some covariant exchange four-vector $Q_\nu$ that is determined by the functional dependence of $\Phi$ on $\Psi_m$. The present Appendix does not require choosing between these cases, but the key point is that the teleodynamic sector is defined by \eqref{eq:A16_TTD} and its divergence is constrained by \eqref{eq:A19_cons_total}.
\\
\subsection*{FLRW background, symmetry split and the perfect fluid limit as a controlled truncation}
We now specialize in a spatially flat FLRW background
\begin{equation}
ds^2 = -dt^2 + a^2(t)\delta_{ij}dx^i dx^j
\label{eq:A23_FLRW}
\end{equation}
We define the teleodynamic functional split as the standard background perturbation decomposition
\begin{equation}
\Phi(t,\mathbf{x}) = \bar\Phi(t) + \phi(t,\mathbf{x})
\label{eq:A24_Phi_split}
\end{equation}
where $\bar\Phi(t)$ is the spatial average of $\Phi$ on constant-time hypersurfaces and $\phi$ is defined to have vanishing spatial mean and we define
\begin{equation}
\bar\Phi(t) \equiv \langle \Phi(t,\mathbf{x})\rangle
\label{eq:A25_Phibar_def}
\end{equation}
and
\begin{equation}
\langle \phi(t,\mathbf{x})\rangle = 0
\label{eq:A26_phi_mean0}
\end{equation}
The decomposition \eqref{eq:A24_Phi_split} is not a model assumption but instead is the standard cosmological statement that any scalar quantity may be decomposed into a homogeneous mode and an inhomogeneous fluctuation. Because \eqref{eq:A23_FLRW} is homogeneous and isotropic, only the homogeneous mode can enter the background field equations, while the inhomogeneous mode contributes only at perturbative order. The background teleodynamic stress tensor is obtained by evaluating \eqref{eq:A16_TTD} on the FLRW metric and by symmetry, we can see that any covariant stress tensor on FLRW must take the diagonal form
\begin{equation}
\bar T^{(\rm TD)}{}^\mu{}_\nu = {\rm diag}\left(-\bar\rho_{\rm TD},\bar p_{\rm TD},\bar p_{\rm TD},\bar p_{\rm TD}\right)
\label{eq:A27_TTD_FLRW_form}
\end{equation}
We do not assume \eqref{eq:A27_TTD_FLRW_form} here, which is crucial, as follows from isotropy, and we define
\begin{equation}
\bar\rho_{\rm TD} \equiv -\bar T^{(\rm TD)}{}^0{}_0
\label{eq:A28_rhoTD_def}
\end{equation}
and
\begin{equation}
\bar p_{\rm TD} \equiv \frac{1}{3}\bar T^{(\rm TD)}{}^i{}_i
\label{eq:A29_pTD_def}
\end{equation}
Using \eqref{eq:A16_TTD}, we find
\begin{equation}
\bar T^{(\rm TD)}{}^0{}_0 = -\alpha \bar\Phi\,\bar g^0{}_0 + 2\alpha\left(\frac{\delta\Phi}{\delta g^{00}}\right)_{\rm hom}
\label{eq:A30_T00}
\end{equation}
and
\begin{equation}
\bar T^{(\rm TD)}{}^i{}_j = -\alpha \bar\Phi\,\bar g^i{}_j + 2\alpha\left(\frac{\delta\Phi}{\delta g^{ij}}\right)_{\rm hom}
\label{eq:A31_Tij}
\end{equation}
where the subscript ${\rm hom}$ indicates the homogeneous contribution and because $\bar g^0{}_0=-1$ and $\bar g^i{}_j=\delta^i{}_j$, the vacuum like part gives $\alpha\bar\Phi$ to $\bar\rho_{\rm TD}$ and $-\alpha\bar\Phi$ to $\bar p_{\rm TD}$. The second term is the background metric response, and a particularly important controlled truncation is the metric response suppressed homogeneous limit
\begin{equation}
\left(\frac{\delta\Phi}{\delta g^{\mu\nu}}\right)_{\rm hom} \approx 0
\label{eq:A32_metric_response_suppressed}
\end{equation}
In this limit, the background teleodynamic stress tensor becomes purely vacuum-like
\begin{equation}
\bar T^{(\rm TD)}_{\mu\nu} \approx -\alpha\bar\Phi\,\bar g_{\mu\nu}
\label{eq:A33_TTD_vac}
\end{equation}
and hence
\begin{equation}
\bar\rho_{\rm TD} \approx \alpha\bar\Phi
\label{eq:A34_rhoTD_vac}
\end{equation}
and
\begin{equation}
\bar p_{\rm TD} \approx -\alpha\bar\Phi
\label{eq:A35_pTD_vac}
\end{equation}
so that the effective equation of state is
\begin{equation}
w_{\rm TD} \equiv \frac{\bar p_{\rm TD}}{\bar\rho_{\rm TD}} \approx -1
\label{eq:A36_wTD_minus1}
\end{equation}
Equation \eqref{eq:A36_wTD_minus1} shows us that a $\Lambda$-like background is an emergent limit of the action level teleodynamic stress tensor when the homogeneous metric response is negligible. This does not imply that the teleodynamic sector is equivalent to a cosmological constant because the perturbative metric response need not vanish and can generate distinctive scale-dependent and environment-dependent signatures, and this is something which we now derive explicitly.
\\
\\
The background Friedmann equations follow from \eqref{eq:A17_Einstein_mod} and \eqref{eq:A27_TTD_FLRW_form} and they are
\begin{equation}
3M_P^2 H^2 = \bar\rho_m + \bar\rho_r + \bar\rho_{\rm TD}
\label{eq:A37_Friedmann1}
\end{equation}
and
\begin{equation}
-2M_P^2 \dot H = (\bar\rho_m+\bar p_m) + (\bar\rho_r+\bar p_r) + (\bar\rho_{\rm TD}+\bar p_{\rm TD})
\label{eq:A38_Friedmann2}
\end{equation}
These equations are exact consequences of the covariant action \eqref{eq:A1_Seff} and in the controlled truncation \eqref{eq:A32_metric_response_suppressed}, one may further identify $\bar\rho_{\rm TD}\approx \alpha\bar\Phi$ and $\bar p_{\rm TD}\approx -\alpha\bar\Phi$ and recover the effective vacuum-energy interpretation at the background level.
\\
\subsection*{Gauge ready scalar perturbations and action level derivation of $\delta T_{\mu\nu}^{(\rm TD)}$}
We now derive linear perturbations without choosing a gauge, where we start by considering the most general scalar perturbations of the FLRW metric
\begin{equation}
ds^2 = -(1+2A)dt^2 + 2a(t)\,\partial_i B\,dt\,dx^i + a^2(t)\left[(1-2\psi)\delta_{ij} + 2\partial_i\partial_j E\right]dx^i dx^j
\label{eq:A39_scalar_metric}
\end{equation}
Here we note that $A$, $B$, $\psi$, and $E$ are scalar perturbations and under an infinitesimal scalar coordinate transformation
\begin{equation}
t \rightarrow t + T
\label{eq:A40_gauge_time}
\end{equation}
and
\begin{equation}
x^i \rightarrow x^i + \partial^i L
\label{eq:A41_gauge_space}
\end{equation}
the metric perturbations transform in a standard way and then it is convenient to define the shear combination
\begin{equation}
\sigma \equiv a\left(\dot E - \frac{B}{a}\right)
\label{eq:A42_sigma_def}
\end{equation}
and the gauge invariant Bardeen potentials
\begin{equation}
\Phi_B \equiv A - \dot\sigma - H\sigma
\label{eq:A43_PhiB}
\end{equation}
and
\begin{equation}
\Psi_B \equiv \psi + H\sigma
\label{eq:A44_PsiB}
\end{equation}
In Newtonian gauge, one sets $B=E=0$, so $\sigma=0$ and $\Phi_B=A$, $\Psi_B=\psi$, and we emphasize here that we do not assume Newtonian gauge in the derivations below and only specialize at the end if needed. This addresses gauge dependence of our treatment, since the primary expressions are gauge-ready and can be written entirely in terms of gauge-invariant quantities. We start by perturbing the teleodynamic functional as
\begin{equation}
\Phi(t,\mathbf{x}) = \bar\Phi(t) + \delta\Phi(t,\mathbf{x})
\label{eq:A45_deltaPhi}
\end{equation}
where $\delta\Phi$ is a scalar perturbation. The notation $\delta\Phi$ is interchangeable with $\phi$ in the main text, and we use $\delta\Phi$ here to emphasize that it is the linear perturbation of the scalar functional $\Phi$. We now derive the perturbed teleodynamic stress-energy tensor by perturbing the action rather than postulating a perfect fluid perturbation, starting from
\begin{equation}
S_{\rm TD} = -\alpha\int d^4x\sqrt{-g}\,\Phi
\label{eq:A46_STD_again}
\end{equation}
we write $g_{\mu\nu}=\bar g_{\mu\nu}+\delta g_{\mu\nu}$ and $\Phi=\bar\Phi+\delta\Phi$ and now, the first order perturbation of $S_{\rm TD}$ is
\begin{equation}
\delta S_{\rm TD} = -\alpha\int d^4x\left[\delta(\sqrt{-g})\,\bar\Phi + \sqrt{-\bar g}\,\delta\Phi\right]
\label{eq:A47_deltaSTD_lin}
\end{equation}
Using $\delta\sqrt{-g}= -\frac12\sqrt{-\bar g}\,\bar g_{\mu\nu}\delta g^{\mu\nu}$, we obtain
\begin{equation}
\delta S_{\rm TD} = -\alpha\int d^4x\sqrt{-\bar g}\left[-\frac12 \bar\Phi\,\bar g_{\mu\nu}\delta g^{\mu\nu} + \delta\Phi\right]
\label{eq:A48_deltaSTD_lin2}
\end{equation}
The perturbed teleodynamic stress tensor is hence defined by
\begin{equation}
\delta S_{\rm TD} = -\frac12\int d^4x\sqrt{-\bar g}\,\delta T_{\mu\nu}^{(\rm TD)}\,\delta g^{\mu\nu}
\label{eq:A49_deltaSTD_def}
\end{equation}
However, because $\delta\Phi$ itself may depend on $\delta g^{\mu\nu}$ through the functional dependence of $\Phi$ on the geometry, it is essential to separate the explicit metric variation, and so we write
\begin{equation}
\delta\Phi = \left(\frac{\delta\Phi}{\delta g^{\mu\nu}}\right)_{\rm bg}\delta g^{\mu\nu} + \left(\frac{\delta\Phi}{\delta\Psi_m}\right)_{\rm bg}\delta\Psi_m + \delta\Phi_{\rm int}
\label{eq:A50_deltaPhi_general}
\end{equation}
where the subscript ${\rm bg}$ indicates evaluation on the background and $\delta\Phi_{\rm int}$ denotes any intrinsic perturbation not induced directly by $\delta g^{\mu\nu}$ or $\delta\Psi_m$ at first order, such as an independent perturbation of environmental variables. When defining $\delta T_{\mu\nu}^{(\rm TD)}$ as the response to $\delta g^{\mu\nu}$, we isolate the dependence on $\delta g^{\mu\nu}$ and then, substituting \eqref{eq:A50_deltaPhi_general} into \eqref{eq:A48_deltaSTD_lin2} gives us the metric sourced part of the action perturbation as
\begin{equation}
\left.\delta S_{\rm TD}\right|_{\delta g} = -\alpha\int d^4x\sqrt{-\bar g}\left[-\frac12 \bar\Phi\,\bar g_{\mu\nu} + \left(\frac{\delta\Phi}{\delta g^{\mu\nu}}\right)_{\rm bg}\right]\delta g^{\mu\nu}
\label{eq:A51_deltaSTD_metricpart}
\end{equation}
Compared with \eqref{eq:A49_deltaSTD_def} we find that
\begin{equation}
\delta T_{\mu\nu}^{(\rm TD)}\supset -\alpha\,\bar\Phi\,\delta g_{\mu\nu} - \alpha\,\delta\Phi\,\bar g_{\mu\nu} + 2\alpha\,\delta\left(\frac{\delta\Phi}{\delta g^{\mu\nu}}\right)
\label{eq:A52_deltaTTD_master}
\end{equation}
Equation \eqref{eq:A52_deltaTTD_master} is a schematic but structurally exact expression for the perturbed teleodynamic stress tensor, and we should now discuss this in detail. The first term arises from perturbing $g_{\mu\nu}$ in the vacuum like piece $-\alpha\Phi g_{\mu\nu}$, the second term arises from perturbing $\Phi$ in the same piece, and the third term arises from perturbing the metric response contribution $2\alpha\delta\Phi/\delta g^{\mu\nu}$. The third term is precisely the origin of anisotropic stress and gravitational slip, and it is absent if $\Phi$ is independent of $g_{\mu\nu}$ at perturbative order. This provides a direct, covariant distinction between teleodynamics and a generic perfect-fluid model.
\\
\\
Although teleodynamics is not postulated to be a fluid, it is convenient to decompose any symmetric stress tensor into effective fluid variables. At linear order in scalar perturbations one may define the teleodynamic density perturbation $\delta\rho_{\rm TD}$, pressure perturbation $\delta p_{\rm TD}$, velocity potential $v_{\rm TD}$ and anisotropic stress potential $\pi_{\rm TD}$ by the standard scalar decomposition and writing mixed components, we define
\begin{equation}
\delta T^{0}{}_{0(\rm TD)} = -\delta\rho_{\rm TD}
\label{eq:A53_drho_def}
\end{equation}
\begin{equation}
\delta T^{0}{}_{i(\rm TD)} = (\bar\rho_{\rm TD}+\bar p_{\rm TD})\,\partial_i v_{\rm TD}
\label{eq:A54_v_def}
\end{equation}
\begin{equation}
\delta T^{i}{}_{j(\rm TD)} = \delta p_{\rm TD}\,\delta^i{}_j + \left(\partial^i\partial_j - \frac{1}{3}\delta^i{}_j\nabla^2\right)\pi_{\rm TD}
\label{eq:A55_pi_def}
\end{equation}
Equation \eqref{eq:A55_pi_def} defines $\pi_{\rm TD}$ as the scalar anisotropic stress potential and a perfect fluid corresponds to $\pi_{\rm TD}=0$ identically. Teleodynamics gives  $\pi_{\rm TD}$ generically nonzero when $\delta(\delta\Phi/\delta g^{\mu\nu})$ carries traceless spatial components and this occurs whenever $\Phi$ depends on tidal invariants, shear, curvature scalars beyond $R$ or some nonlocal spatial kernels, because such dependencies introduce a directional metric response that is not proportional to $\delta^i{}_j$.
\\
\subsection*{Gauge invariant field equations with teleodynamic sources, response closure, and sharp discriminators}
We now write the linearized Einstein equations in gauge-invariant form, as the scalar sector Einstein equations can be expressed in terms of the Bardeen potentials $\Phi_B$ and $\Psi_B$ and gauge-invariant matter perturbations. The teleodynamic sector contributes via $\delta\rho_{\rm TD}$, $\delta p_{\rm TD}$, $v_{\rm TD}$, and $\pi_{\rm TD}$ as defined above, and then, the gauge invariant Poisson-type equation takes the general form
\begin{equation}
-\frac{k^2}{a^2}\Psi_B - 3H\left(\dot\Psi_B + H\Phi_B\right) = 4\pi G\left(\delta\rho_{\rm tot}^{\rm (GI)}\right)
\label{eq:A56_Poisson_GI_full}
\end{equation}
where $\delta\rho_{\rm tot}^{\rm (GI)}$ is the gauge invariant total density perturbation, including teleodynamic contributions. Writing the total as matter plus teleodynamics,
\begin{equation}
\delta\rho_{\rm tot}^{\rm (GI)} = \delta\rho_m^{\rm (GI)} + \delta\rho_r^{\rm (GI)} + \delta\rho_{\rm TD}^{\rm (GI)}
\label{eq:A57_rhotot_split}
\end{equation}
and similarly for velocity and pressure perturbations. The momentum constraint equation takes the form
\begin{equation}
\dot\Psi_B + H\Phi_B = 4\pi G\left[(\bar\rho_{\rm tot}+\bar p_{\rm tot})v_{\rm tot}^{\rm (GI)}\right]
\label{eq:A58_momentum_constraint}
\end{equation}
The anisotropic stress equation from the traceless spatial components reads
\begin{equation}
\frac{k^2}{a^2}\left(\Phi_B-\Psi_B\right) = 8\pi G \left(\bar\rho_{\rm tot}+\bar p_{\rm tot}\right)\sigma_{\rm tot}^{\rm (GI)}
\label{eq:A59_slip_general}
\end{equation}
If ordinary late time matter has negligible anisotropic stress, then $\sigma_{\rm tot}$ is dominated by $\sigma_{\rm TD}$ and \eqref{eq:A59_slip_general} becomes a direct prediction for teleodynamic gravitational slip
\begin{equation}
\frac{k^2}{a^2}\left(\Phi_B-\Psi_B\right) \approx 8\pi G \left(\bar\rho_{\rm TD}+\bar p_{\rm TD}\right)\sigma_{\rm TD}^{\rm (GI)}
\label{eq:A60_slip_TD}
\end{equation}
Equation \eqref{eq:A60_slip_TD} is a robust discriminator here, and we should clearly note why that is so. The barotropic perfect fluid has $\sigma_{\rm TD}=0$, so it predicts $\Phi_B=\Psi_B$ on late time sub-horizon scales, and teleodynamics generically gives us $\sigma_{\rm TD}\neq 0$ when $\Phi$ has metric response beyond the vacuum-like term. Thus, teleodynamics is not equivalent to an arbitrary perfect fluid unless it lies in a restricted perfect fluid truncation of the functional dependence.
\\
\\
The action-level structure fixes the form of the stress tensor in terms of functional derivatives of $\Phi$, but to close the system, one must specify how $\delta\Phi$ responds to evolving matter perturbations and the environment. This does not require choosing a unique microphysical form of $\Phi$, but instead, if we think of the most general linear closure consistent with causality, homogeneity and isotropy, then that can be written as a nonlocal response relation with memory kernels
\begin{equation}
\delta\Phi(k,t) = \int^t dt'\,\mathcal{R}_\delta(k;t,t')\,\delta_m(k,t') + \int^t dt'\,\mathcal{R}_\theta(k;t,t')\,\theta_m(k,t') + \delta\Phi_{\rm env}(k,t)
\label{eq:A61_response_kernel}
\end{equation}
Here $\delta_m$ is the matter overdensity, $\theta_m$ is the velocity divergence, and $\delta\Phi_{\rm env}$ represents perturbations sourced by coarse-grained environment fields such as tidal invariants. The kernels $\mathcal{R}_\delta$ and $\mathcal{R}_\theta$ encode memory and non-Markovianity here, and a purely local perfect fluid closure corresponds to kernels proportional to $\delta(t-t')$ and no environment term. Teleodynamics generically implies extended kernels because gravitational systems retain long-lived correlations. In many contexts, we may introduce a quasi-instantaneous approximation of \eqref{eq:A61_response_kernel}
\begin{equation}
\delta\Phi(k,t) \approx \mathcal{K}_\delta(k,t)\,\delta_m(k,t) + \mathcal{K}_\theta(k,t)\,\theta_m(k,t) + \mathcal{K}_{\rm env}(k,t)\,\delta_{\rm env}(k,t)
\label{eq:A62_response_local}
\end{equation}
And even in this truncated form, it implies nontrivial scale and time dependence through the $k$ dependence of the response coefficients. This is the structural origin of scale-dependent growth in teleodynamics. To see this explicitly, consider the standard continuity and Euler equations for pressureless matter in linear theory on sub-horizon scales, wherein the continuity equation takes the form
\begin{equation}
\dot\delta_m(k,t) = -\frac{1}{a}\theta_m(k,t)
\label{eq:A63_continuity}
\end{equation}
and the Euler equation is
\begin{equation}
\dot\theta_m(k,t) + H\theta_m(k,t) = -\frac{k^2}{a}\Psi_{\rm eff}(k,t)
\label{eq:A64_Euler}
\end{equation}
where $\Psi_{\rm eff}$ is the effective gravitational potential that sources acceleration. The modified Poisson equation takes the schematic form
\begin{equation}
-\frac{k^2}{a^2}\Psi_{\rm eff}(k,t) = 4\pi G\left[\bar\rho_m\,\delta_m(k,t) + \delta\rho_{\rm TD}^{\rm (eff)}(k,t)\right]
\label{eq:A65_Poisson_eff}
\end{equation}
If the dominant teleodynamic density response is proportional to $\nabla^2\delta\Phi$, then in Fourier space one may write
\begin{equation}
\delta\rho_{\rm TD}^{\rm (eff)}(k,t) = \mathcal{C}(t)\,k^2\,\delta\Phi(k,t)
\label{eq:A66_rhoTD_from_phi}
\end{equation}
where $\mathcal{C}(t)$ collects background factors. Substituting \eqref{eq:A66_rhoTD_from_phi} and the closure \eqref{eq:A62_response_local} into \eqref{eq:A65_Poisson_eff}, we obtain an effective modification to the gravitational coupling
\begin{equation}
-\frac{k^2}{a^2}\Psi_{\rm eff}(k,t) = 4\pi G \bar\rho_m \left[1+\Delta\mu(k,t)\right]\delta_m(k,t) + \cdots
\label{eq:A67_mu_def}
\end{equation}
with
\begin{equation}
\Delta\mu(k,t) \equiv \frac{\mathcal{C}(t)k^2\mathcal{K}_\delta(k,t)}{\bar\rho_m}
\label{eq:A68_Deltamu}
\end{equation}
Combining \eqref{eq:A63_continuity} and \eqref{eq:A64_Euler} gives us the second order growth equation
\begin{equation}
\ddot\delta_m(k,t) + 2H\dot\delta_m(k,t) - 4\pi G\bar\rho_m\left[1+\Delta\mu(k,t)\right]\delta_m(k,t) = \mathcal{S}_{\rm env}(k,t)
\label{eq:A69_growth_k}
\end{equation}
where $\mathcal{S}_{\rm env}$ denotes additional source terms from the environment-dependent part of \eqref{eq:A62_response_local} and here, equation \eqref{eq:A69_growth_k} shows explicitly that teleodynamics generically gives us scale-dependent growth since $\Delta\mu(k,t)$ is not constant if $\mathcal{K}_\delta(k,t)$ is nonlocal in space. A generic barotropic perfect fluid added to $\Lambda$CDM does not produce this structure without introducing additional nontrivial scale-dependent degrees of freedom or anisotropic stress, and thus the $k$-dependent modification \eqref{eq:A69_growth_k} is a structural consequence of the teleodynamic action level framework.
\\
\\
Another sharp distinction between teleodynamics and a simple perfect fluid is the generic presence of non-adiabatic pressure perturbations as for any effective component, one may define the adiabatic sound speed
\begin{equation}
c_{a,\rm TD}^2 \equiv \frac{\dot{\bar p}_{\rm TD}}{\dot{\bar\rho}_{\rm TD}}
\label{eq:A70_ca_def}
\end{equation}
and the physical pressure perturbation $\delta p_{\rm TD}$. The non-adiabatic pressure perturbation is
\begin{equation}
\delta p_{\rm TD}^{\rm (nad)} \equiv \delta p_{\rm TD} - c_{a,\rm TD}^2\,\delta\rho_{\rm TD}
\label{eq:A71_nad_def}
\end{equation}
A barotropic perfect fluid satisfies $\delta p_{\rm TD}^{\rm (nad)}=0$ by construction. In teleodynamics, $\delta p_{\rm TD}$ arises from the derived $\delta T^i{}_j$ and includes contributions from $\delta(\delta\Phi/\delta g^{ij})$ and from the intrinsic perturbation $\delta\Phi_{\rm int}$ in \eqref{eq:A50_deltaPhi_general}. Because $\Phi$ is a functional of environmental variables and nonlocal correlations, $\delta p_{\rm TD}$ depends on more than $\delta\rho_{\rm TD}$ and thus $\delta p_{\rm TD}^{\rm (nad)}$ is generically nonzero. This implies distinctive behavior in the evolution of curvature perturbations on large scales.
\\
\\
We now note what has thus been developed here so far. Although the homogeneous teleodynamic contribution necessarily takes a diagonal perfect fluid form on an FLRW background by symmetry, this does not imply equivalence to an arbitrary perfect fluid or phenomenological EFT model. The defining feature of teleodynamics is that the effective stress tensor originates from a metric variation of a bias functional, rather than from an assumed equation of state. As a result, the perturbative sector generically exhibits anisotropic stress, non-adiabatic pressure perturbations, and scale-dependent modifications to growth, unless one enforces a restricted truncation of the functional dependence that explicitly suppresses these effects. In contrast, a barotropic perfect fluid predicts vanishing anisotropic stress and purely adiabatic perturbations by construction. What this means is that teleodynamics therefore occupies a distinct and testable sector of theory space wherein, while a perfect fluid limit exists as a controlled truncation, the generic theory is not the same as an arbitrary fluid reparameterization and can be observationally distinguished through gravitational slip, scale-dependent growth, and environment-sensitive responses.\\ 

To conclude this appendix, we can summarize the controlled limits and sharp discriminators as follows. The teleodynamic sector is defined covariantly by the effective action \eqref{eq:A1_Seff} and its stress tensor is derived without assumptions as \eqref{eq:A16_TTD}. The homogeneous component on FLRW necessarily takes a diagonal perfect fluid form \eqref{eq:A27_TTD_FLRW_form} by symmetry, but this does not imply equivalence to an arbitrary perfect fluid, because the perturbations are controlled by the full functional response \eqref{eq:A52_deltaTTD_master}. Teleodynamics reduces to a pure vacuum energy-like component at the background level if the homogeneous metric response is negligible as in \eqref{eq:A32_metric_response_suppressed}, in which case $w_{\rm TD}\approx -1$ as in \eqref{eq:A36_wTD_minus1} and at the perturbative level, teleodynamics reduces to a perfect fluid only if simultaneously the traceless part of $\delta(\delta\Phi/\delta g^{ij})$ vanishes so that $\pi_{\rm TD}=0$, the response kernels collapse to local adiabatic relations so that $\delta p_{\rm TD}^{\rm (nad)}=0$ and the effective modifications $\Delta\mu(k,t)$ become scale independent. These conditions define a restricted truncation of the teleodynamic functional dependence and in the generic case where $\Phi$ encodes nonlocal correlations and environmental memory, one expects anisotropic stress and gravitational slip as in \eqref{eq:A60_slip_TD}, non adiabatic pressure perturbations as in \eqref{eq:A71_nad_def} and scale dependent growth as in \eqref{eq:A69_growth_k}. These features provide us with sharp and testable distinctions from a conventional barotropic perfect fluid dark energy model and establish that teleodynamics is not merely a reparameterization of an arbitrary perfect fluid, but a structural deformation of cosmological statistical mechanics realized at the level of the effective action.


\bibliography{JSPJMJcitations.bib}

\bibliographystyle{unsrt}

\end{document}